\documentclass[preprint,3p,times,sort&compress,nopreprintline]{elsarticle}

\usepackage{amsmath,amssymb}
\usepackage{graphicx}
\usepackage{float}
\usepackage{xcolor}
\usepackage{subcaption}
\usepackage{esint} 
\usepackage{hyperref}
\usepackage{orcidlink}
\usepackage{sectsty}
\allsectionsfont{\bfseries}
\usepackage{fancyhdr}
\makeatletter
\patchcmd{\pprintMaketitle}{\footnotesize\itshape\elsaddress}{\footnotesize\normalfont\elsaddress}{}{}
\patchcmd{\MaketitleBox}{\footnotesize\itshape\elsaddress}{\footnotesize\normalfont\elsaddress}{}{}
\makeatother
\begin{document}

\title{Shimmer: End-to-End Open-Source Passive $\mathbf{B}_0$ Shimming Methodology for Low-Field MRI Magnets}
\author[1]{Ilia Kulikov\orcidlink{0000-0002-7544-2124}}
\author[1]{David Schote\orcidlink{0000-0003-3468-0676}}
\author[2]{Helge Herthum\orcidlink{0000-0001-6494-0833}}
\author[3,5]{Martin Häuer\orcidlink{0000-0002-1611-7129}}
\author[1]{Kimon Hadjikiriakos\orcidlink{0009-0002-8625-8522}}
\author[1]{Jan Gregor Frintz\orcidlink{0009-0000-9389-4178}}
\author[4]{Sonja Buxbaum-Conradi\orcidlink{0009-0007-1925-9556}}
\author[4]{Manuel Moritz\orcidlink{0000-0001-5126-9016}}
\author[4]{Tobias Redlich\orcidlink{0000-0003-4129-8926}}
\author[1,5]{Lukas Winter\orcidlink{0000-0002-4381-275X}}

\affiliation[1]{organization={Physikalisch-Technische Bundesanstalt (PTB)}, addressline={8.1 Biomedical Magnetic Resonance, Abbestr. 2-12 10587}, city={Berlin}, country={Germany}}

\affiliation[2]{organization={Berlin Center for Advanced Neuroimaging, Charite--Universitatsmedizin Berlin, Corporate Member of Freie Universitat Berlin, Berlin Institute of Health}, city={Berlin}, country={Germany}}

\affiliation[3]{organization={Just Transition Center, Group C1, Martin-Luther-Universität Halle-Wittenberg}, city={Halle (Saale)}, country={Germany}}

\affiliation[4]{organization={Helmut Schmidt University}, city={Hamburg}, country={Germany}}

\affiliation[5]{organization={Open Source Imaging Initiative (OSI²) e.V.}, city={Berlin}, country={Germany}}

\cortext[cor]{Corresponding author.}
\ead{lukas.winter@ptb.de}

\begin{abstract}
Low-field MRI scanners based on permanent magnet arrays require accurate $\mathbf{B}_0$ shimming to achieve sufficient field homogeneity for imaging. Here, we present Shimmer, an open-source methodology for passive shimming of low-field MRI magnets using additional NdFeB magnets placed outside the main magnet array. The end-to-end methodology combines magnetic field mapping, magnetic field simulations, numerical optimization of shim magnet positions and orientations, and automated generation of 3D-printable shim holders. Continuous magnet rotations and several optimization strategies are considered and evaluated. The method was validated and applied to three different permanent magnet arrays ($\approx$50~mT) in 200~mm diameter spherical volumes, reducing initial field inhomogeneities of 9361~ppm, 31721~ppm and 34143~ppm to 888~ppm, 1812~ppm and 1602~ppm respectively. The improved homogeneity enabled undistorted MR imaging of a phantom. Shimmer provides a reproducible and adaptable approach for improving the performance of low-field MRI magnets.
\end{abstract}


\maketitle


\renewcommand\thefootnote{}
\footnotetext{\textbf{Abbreviations:} DSV - diameter of the spherical volume, OSI$^2$ - Open Source Imaging Initiative, SH - Spherical Harmonic, NN - Neural network}

\renewcommand\thefootnote{\fnsymbol{footnote}}
\setcounter{footnote}{1}

\section{Introduction}\label{intro}

Portable MRI scanners based on Halbach-like arrays of permanent magnets show remarkable imaging performance in point-of-care applications~\cite{winterOpensourceMagneticResonance2024, schoteherthum2026, guallart-navalPortableMagneticResonance2022, mcdanielMRCapSinglesided2019,obungolochOnsiteConstructionPointofcare2023,galveEllipticalHalbachMagnet2024,schoteNexusVersatileConsole2025a,rodriguezOpenSourceMultinuclearLowField2026}. Combined with open-source software and open-source hardware, these scanners have already developed into an innovation platform for academic research globally and with first ongoing commercial adaptations, they hold the promise to become an affordable diagnostic tool for science and healthcare. In contrast to traditional higher-field ($\mathbf{B}_0>0.5$~T) MRI systems that are based on superconducting magnets, compact ultralow-field ($\mathbf{B}_0\approx50$~mT) magnets consist of $>1000$ individual small permanent magnets that do not require high power supplies and cryogenic cooling. However, one of the key challenges of constructing these magnets is the $\mathbf{B}_0$ homogeneity of the magnetic field inside the target imaging volume~\cite{cooleyTwodimensionalImagingLightweight2015,cooleyDesignSparseHalbach2018,oreillyThreedimensionalMRIHomogenous2019,tewariPermanentMagnetHypothesis2023,blockMRI4ALLWeekLongHackathon2025,kleinFerrofluidsImproveField2022,galveEllipticalHalbachMagnet2024}. 
Though simulations may predict highly homogeneous $\mathbf{B}_0$ fields~\cite{cooleyTwodimensionalImagingLightweight2015,cooleyDesignSparseHalbach2018}, deviations of the position, rotation or remanence of the many individual magnets used may lead to mismatches in the order of several magnitudes between simulated and constructed magnets~\cite{zanovelloVeryLowFieldMRIScanners2025}. 
These discrepancies lead to very inhomogeneous magnetic fields making useful imaging applications challenging or even impossible for several reasons. The RF coils used at 50~mT ($f_0$=2.1~MHz) are typically single channel transmit receive coils with a high Q factor to achieve high signal-to-noise ratios (SNR)~\cite{vliemDesignPerformanceToroidal2026}. For an S11 -3~dB bandwidth of the RF coil, values around 11~kHz can be reached~\cite{webbTacklingSNRLowfield2023}, which translates to 5170~ppm considering a 50~mT magnet. To effectively excite all spins and maintain flip angle and receive signal fidelity within the complete imaging volume and keep image distortions manageable, a good target for $\mathbf{B}_0$ field inhomogeneity is around or below 1000~ppm in the target field-of-view. Within these inhomogeneities, high quality in-vivo images of the human brain have been recorded ~\cite{oreillyVivo3DBrain2021,zhaoWholebodyMagneticResonance2024,schoteNexusVersatileConsole2025a}
In clinical high-field systems based on superconducting magnets, the initial field inhomogeneity of the constructed magnets is in the order of hundreds of~ppm~\cite{kongNovelPassiveShimming2015}, an order of magnitude lower than for the initially constructed Halbach arrays, and those field distortions are predominantly due to low spherical harmonics~\cite{abePassiveShimmingMRI2017}. Consequently, the superconducting magnet can be effectively shimmed with iron blocks that redirect the $\mathbf{B}_0$ and effectively correct for the low-order terms ~\cite{abePassiveShimmingMRI2017}, ultimately resulting in a $\mathbf{B}_0$ field inhomogeneity below 5~ppm~\cite{kongNovelPassiveShimming2015}. For lower-cost low-field Halbach magnets, the field distortions of the initially constructed magnets are much higher, containing stronger contributions from higher spherical harmonics and provide less inner bore space ~\cite{wenzelB0ShimmingMethodologyAffordable2021,cooleyDesignSparseHalbach2018} requiring different strategies to improve field homogeneity. One approach is to install additional shim magnets around the main array~\cite{wenzelB0ShimmingMethodologyAffordable2021,blockMRI4ALLWeekLongHackathon2025}. For that, a map of the magnetic field inside the array is recorded, then the positions and rotations of the shim magnets are computed and the additional magnets (shims) are installed around the main Halbach array. This approach was shown to effectively improve the field homogeneity in cylindrical Halbach arrays~\cite{cooleyDesignSparseHalbach2018,oreillyVivo3DBrain2021}.\\
When the field distortions contain only lower (up to $n\leq2$) spherical harmonics, active $\mathbf{B}_0$ shimming can be applied with a set of active shimming coils to effectively decrease the field inhomogeneity down to a few ppm~\cite{wuShimCoilDesign2018,pinhomenesesShimCoilsTailored2022,juchemDynamicMulticoilShimming2011}. Therefore, the passive shimming procedure is targeted at removing the higher-order spherical harmonic distortions of the field to consecutively apply different, dynamic $\mathbf{B}_0$ shimming approaches.\\

\par
To date, passive shimming procedures for low-field MRI magnets have been reported with
varying levels of methodological detail and mostly missing open-source procedures that can be reproduced, adopted and evaluated. In most cases shimming is peripheral to the main content of the paper rather than its focus. O'Reilly et al.~\cite{oreillyVivo3DBrain2021} shimmed a 50~mT, 27~cm bore Halbach array by optimizing, with a genetic algorithm, the filling of a grid of 900 candidate positions (each left empty or filled with a magnet in one of two orientations), decreasing field inhomogeneity from 13000~ppm to approximately 2400~ppm over a 20~cm DSV (diameter of the spherical volume). This procedure was described only briefly. Galve et al.~\cite{galveEllipticalHalbachMagnet2024} presented and compared different optimization strategies for passive shimming of an elliptical Halbach MRI magnet. While these evaluations are useful, a substantial gap between predicted and measured field inhomogeneity after shimming (3600~ppm vs. 5700~ppm in a head-size volume) is reported, and the final target value, even in simulations, deviates from the envisioned inhomogeneity by around 1000~ppm, leaving some open questions around the efficiency of that approach.\\ 

\par
Cooley et al.~\cite{cooleyPortableScannerMagnetic2020} optimized 2016 shim magnet locations for single sized 6.35~mm NdFeB cubes using a target-field shimming iteration. The method was applied to improve a built-in encoding field gradient and not to improve $\mathbf{B}_0$ field homogeneity, while further information and a detailed evaluation on its effectiveness is missing. Other approaches, like Wenzel et al.~\cite{wenzelB0ShimmingMethodologyAffordable2021} that also share open-source code for their genetic algorithm approach, or Wang et al.~\cite{wangPassiveShimmingMethod2022} who used ferromagnetic sheet arrays, described methods that were applied to very small target volumes for desktop MR magnets or NMR-spectrometer samples, so their applicability for larger target volumes remains questionable. No study reported a complete and open-source end-to-end methodology that addresses the field mapping, optimization, assembly and validation, leaving space for ad hoc choices and errors when the approach is adapted to a new magnet build.\\

\par
Here we present the comprehensive open-source software and methodology Shimmer for improving the homogeneity of a target $\mathbf{B}_0$ field by orders of magnitude by placing small cubic permanent magnets outside the main magnet array using continuous magnet rotations. The Shimmer methodology was successfully applied to improve $\mathbf{B}_0$ field homogeneity of three different builds of OSI$^2$ ONE, an open-source, permanent low-field MR magnet~\cite{osii_magnet_gitlab}. Shimmer integrates various strategies to optimize magnet placement, simulations of the magnetic fields and exports labeled 3D printed structures for placing the shim magnets in the shim trays. The different strategies were evaluated and the improvements in performance were assessed with field mapping and MR imaging on an OSI$^2$ ONE low-field MRI scanner. Shimmer is an open-source project with the source code and files publicly available on \href{https://gitlab.com/osii/shimming/shimmer}{\texttt{gitlab.com/osii/shimming/shimmer}}~\cite{kulikovShimmerPythonBasedOpenSource2026}.

\section{Materials and Methods}\label{methods}

\subsection{Shimming Workflow}
\begin{figure}[ht!]
\centerline{\includegraphics[width=\textwidth]{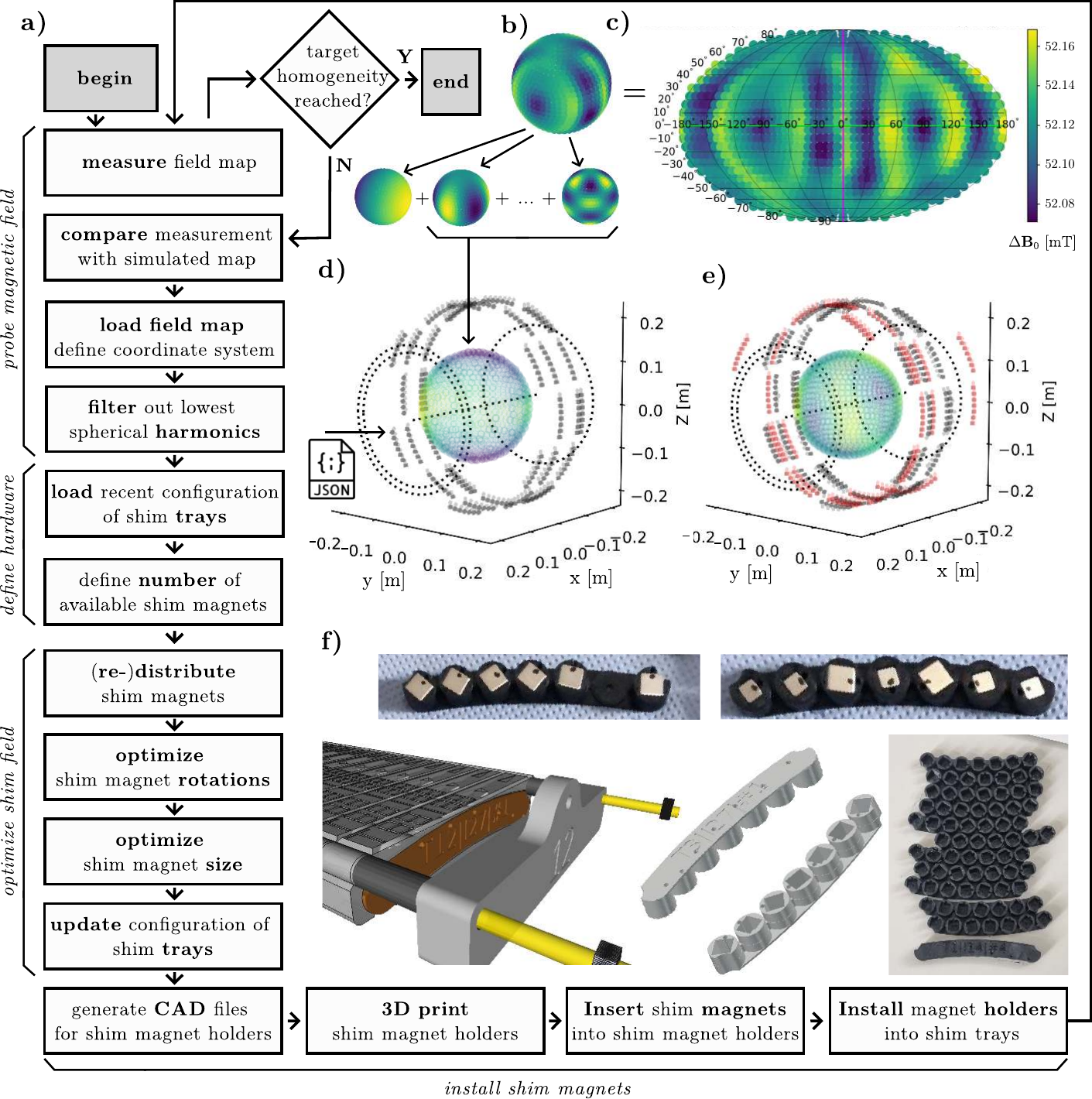}}
\caption{a): flow chart of the iterative passive $\mathbf{B}_0$ shimming process using Shimmer. After probing of the $\mathbf{B}_0$ field, a procedure for placing the shim magnets and adjusting their positions and rotations is performed. The recorded field map is represented as a linear combination of diverging spherical harmonics (b, c, Section~\ref{SI_SH}). The lowest orders are removed. The field map and the current configuration of shims is loaded in the simulated magnet (d). The optimization algorithm places additional shim magnets to the existing shim configuration and improves the field homogeneity (e). 3D printable shim magnet holders are then produced and populated with permanent magnets (f). Finally, the magnet holders are installed into the corresponding addresses. The process repeats until the desired homogeneity is reached.\label{fig_pipeline}}
\end{figure}

A flow chart for the iterative passive shimming process using Shimmer is shown in Figure~\ref{fig_pipeline}. First, the magnetic field inside the magnet bore (Figure~\ref{fig_mollweide_cuts}~a) is measured and the field map is loaded. Previous configurations (for iteration no. $>1$) of shim magnets are loaded over a JSON file. After defining the number of available shim magnet locations for the consecutive $\mathbf{B}_0$ field optimizations, the magnetic fields of the shim magnets at these defined locations (orientation along the main field) are computed at the points where the field map was probed (surface of the DSV). Using optimization algorithms described in Section~\ref{sec:optimizations}, the positions and the rotations of the shim magnets are optimized to decrese the inhomogeneity of the field map. This optimization is repeated for various available shim magnet sizes until the best combination of placement, rotations and size is found. Initially, the positioning of the shim magnets is optimized followed by an optimization of the magnet rotations at each position. After a target homogeneity is reached, the coordinates and rotations of the magnets are used to automatically generate a set of STL models of the shim inserts (see Section~\ref{sec:shim_inserts}), that are used for 3D printing. The shim magnets are inserted in the corresponding shim trays and the current shim configuration is updated in the JSON file. The field map of the magnet with the new shim configuration is measured and compared to the simulations to detect potential errors in the placement of the shim magnets. If needed, the procedure is repeated until the target homogeneity is reached within the specified volume.\\

\subsection{MR Magnets and Passive Shim Inserts}
\label{sec:shim_inserts}
\begin{figure}[h]
\centerline{\includegraphics[width=\textwidth]{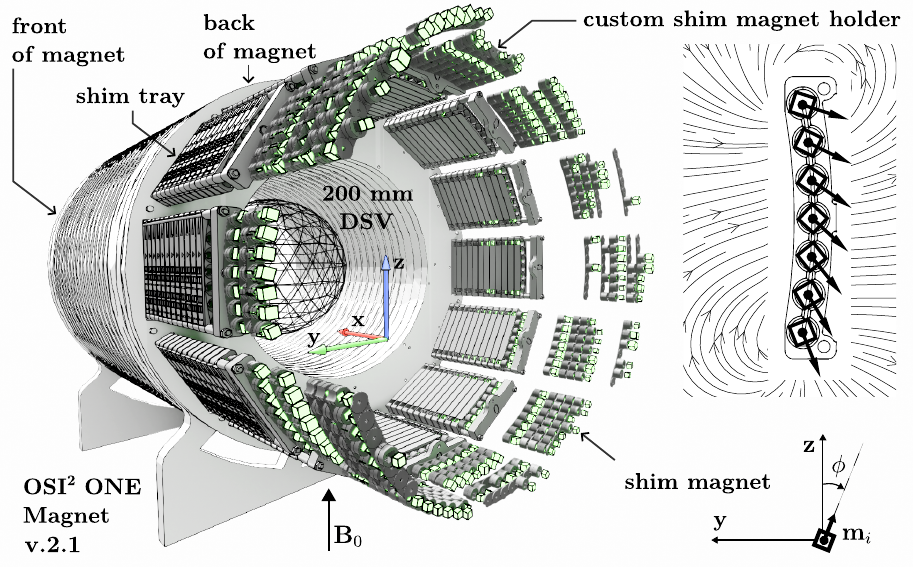}}
\caption{Exemplified assembly of a shimming configuration on a 50~mT OSI$^2$ ONE v2.2 low-field MRI magnet, containing 2320 NdFeB permanent magnets (PM). Probed spherical volume (DSV, 200~mm) in the magnet center. Exploded view of 12 shim trays, containing a custom configuration of 755 magnets of different size and orientations. Each shim tray carries 30 generic shim inserts, each shim insert carries up to 7 shim magnets. The magnets are fixated in the shim insert using a custom 3D printed shim magnet holder. Field lines of one custom magnet holder populated with 7 8-mm magnets is shown on the right. The magnet coordinate system is shown inside of the magnet. The rotation of the dipole moment $\mathbf{m}_i$ of an individual shim magnet is shown as $\phi$ in the magnet coordinate system, counting from the $\mathbf{z}$ axis towards the $-\mathbf{y}$ axis. The direction of the principal component of the static magnetic field $\mathbf{B}_0$ is along $+\mathbf{z}$.\label{fig:magnet_with_shim_inserts}}
\end{figure}

Shimmer was applied to three different low-field MR magnets, the OSI$^2$ ONE v1.0, v2.0 and v2.1~\cite{osii_magnet_gitlab}. These low-field MR magnets consist of Halbach based arrays of small permanent magnets. 
The simulated $\mathbf{B}_0$ inhomogeneity of these magnets is typically an order of magnitude lower than the constructed one. For the OSI$^2$ ONE v2.0 magnet e.g. the simulated $\mathbf{B}_0$ field results in 8460~ppm in a 240~mm DSV and 408~ppm in a 200 mm DSV (Supplementary Figure~\ref{fig_SI_expected_homos}). In comparison, the constructed magnet showed an initial $\mathbf{B}_0$ inhomogeneity of 56001~ppm for a 240~mm sphere and 34143~ppm for a 200~mm sphere, due to build and material tolerances~\cite{zanovelloVeryLowFieldMRIScanners2025}. 
For $\mathbf{B}_0$-shimming, cubic Nd$_2$Fe$_{14}$B magnets with varying size (6~mm, 8~mm and 9~mm) and various grade (N42 with $B_{rem}=1.28-1.32$~T, N45 with $B_{rem}=1.32-1.38$~T and N52 with $B_{rem}=1.43-1.48$~T, at T = 293~K) were inserted in shim inserts (Figure~\ref{fig:magnet_with_shim_inserts}) after their positions and rotations had been optimized. The temperature dependence of the remanence field of these magnets was assumed to be linear with $\alpha = 1200 $~ppm/K and reversible within the used temperature range~ \cite{cooleyDesignSparseHalbach2018,miyamotoDEVELOPMENTPERMANENTMAGNET1989,chenHysteresisModelBased2018}. The change of the $B_{rem}$ with time (10~ppm/yr.~\cite{miyamotoDEVELOPMENTPERMANENTMAGNET1989}) was considered negligible and outliers in $B_{rem}$ due to e.g. production uncertainties were not considered for the optimizations and shim magnet placement~\cite{miyamotoDEVELOPMENTPERMANENTMAGNET1989,wenzelB0ShimmingMethodologyAffordable2021}.\\
\par
\begin{figure}[h]
\centerline{\includegraphics[width=1\textwidth]{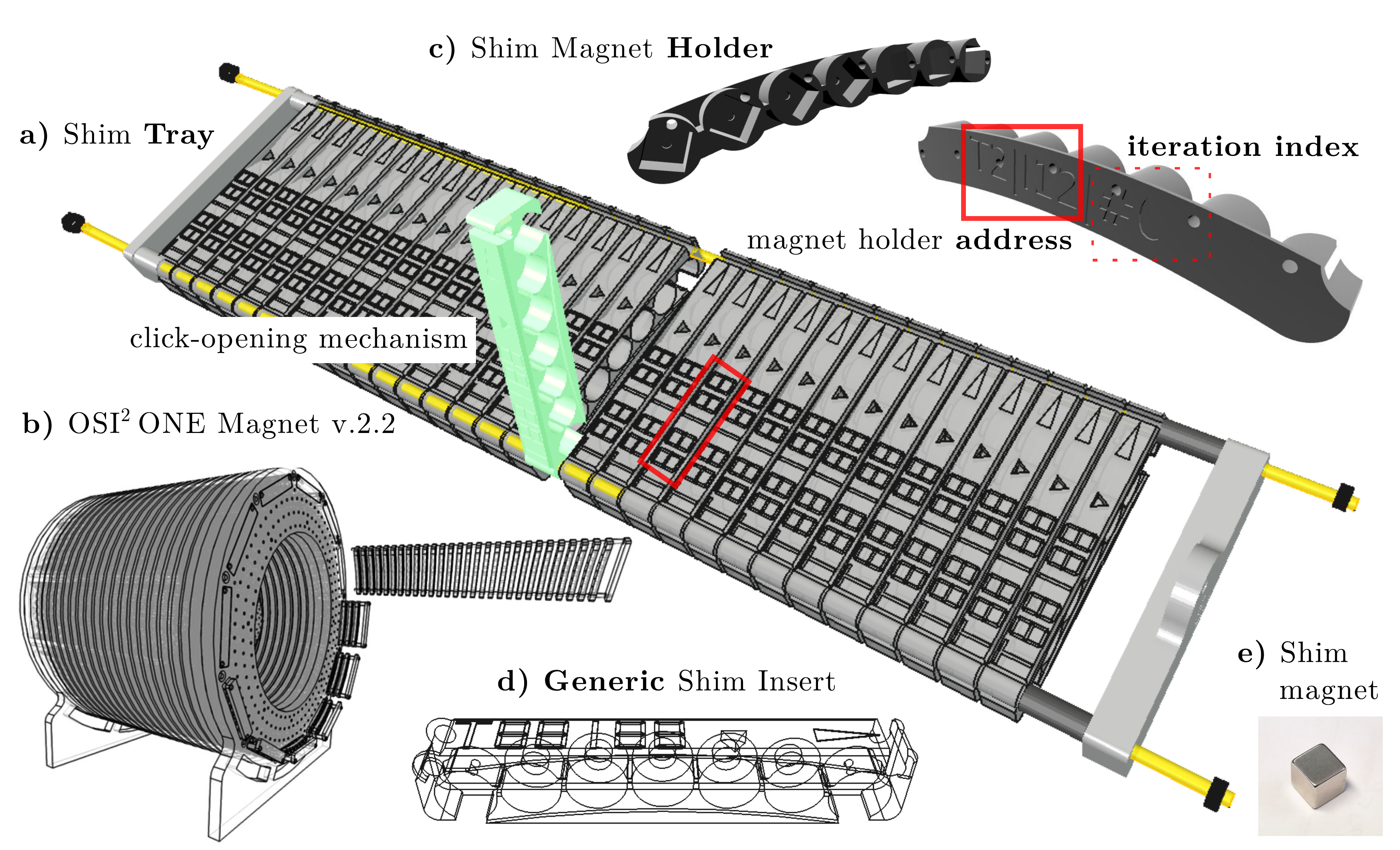}}
\caption{a) Shim tray containing 30 generic shim inserts, b): OSI$^2$ ONE Halbach array with 12 shim trays, c): Custom shim magnet holder, generated by the field optimization script. North pole of a shim magnet is designated with a circular marker, magnet holder address in shim tray (tray $x$ insert $y$, iteration index $z$) is engraved on the back of each magnet holder. d): one generic shim insert, e): 9 mm N42 NdFeB shim magnet.\label{fig_SI_blueprints}}
\end{figure}

A total of 12 shim trays can be placed outside the main magnet as shown in  Figure~\ref{fig:magnet_with_shim_inserts}. To facilitate shimming magnet assembly/disassembly and reduce 3D printing time, a generic shim insert has been designed and a total of 30 \textit{generic shim inserts} can be placed into a shim tray (Figure~\ref{fig_SI_blueprints}~a). Each shim insert can accommodate up to 7 shim magnets for a cubic magnet size up to 9~mm (Figure~\ref{fig_SI_blueprints}~d,e). The shim trays can carry up to 2520 magnets. Locating the shim trays outside the main magnet, allows for low magnetic field forces due to the Halbach design, which enables an easy insertion of the shim trays, and allows for more space for the gradient coils and RF shield inside the magnet bore. Each generic shim insert is attached to brass rods and can be populated with shim magnets individually. The shim inserts consist of two separate parts: 1) a generic part which is attached to the brass rods (Figure~\ref{fig_SI_blueprints}~a,d) and 2) a custom \textit{shim magnet holder} (Figure~\ref{fig_SI_blueprints}~c) that fits into the generic shim insert, which accommodates the rotations needed for a particular magnet size and shim iteration. The generic shim insert was designed in FreeCAD~\cite{freecadFreeCADYourOwn2025} and it can be prefabricated. The custom shim magnet holders are generated for each shim configuration separately. Since the generic shim insert is pre-fabricated, it requires less material to manufacture the custom shim magnet holders reducing 3D printing times substantially. 128 custom shim magnet holders were printed in $<2$ days using $<700$ g of PETG filament. 3D Printing of one generic shim insert took 36 min and 3D printing of one custom shim magnet holder took 33 min using a Bambulab H2D Pro. The STL files of the shim magnet holders are directly generated from the output of the simulation using the trimesh~\cite{dawson-haggertyTrimesh2019} library for Python and they are automatically labeled according to their position in the shim tray with the address (tray|insert, e.g. T2|I12 corresponds to the tray 2, insert 12) to avoid mistakes in magnet placement. The shim magnet holders can be removed easily to change the shim configuration or reuse the shim magnets if desired. 
Populating 12 shim inserts at the same distance along the bore (x-axis) within each shim tray, allows one to create a shim ring at this particular location (Figure~\ref{fig:magnet_with_shim_inserts}).\\ 

\subsection{$\mathbf{B}_0$ Shimming Optimizations}
\label{sec:optimizations}

Optimization of the shim field is a computationally expensive task as many combinations of magnets have to be considered and the optimization of the field is a multidimensional problem. Since the procedure is developed for an arbitrary magnet build, the initial distortions of the field map can vary drastically, so a solid numerical solution of the field optimization problem is required. 
There are several strategies to approach this; one could run a lengthy optimization using all available magnet positions in order to find the configuration with the best possible $\mathbf{B}_0$ field homogeneity. However, since deviations in the actual construction of the shims due to material imperfections and positioning errors will always be present, an iterative approach of fast computations and step-by-step improvements of the magnetic field is applied.

Shimmer aims at using the least amount of shim magnets for the current shim iteration, to enable the most free shim magnet locations for a possible consecutive shim iteration. Different optimization strategies were evaluated that include variations of position, rotation and the size of the shim magnets. 
Depending on the used algorithm, different criteria of the shim field quality (cost functions) are applied for optimizing the shim configuration.\\
\par

\subsubsection{Magnetic Field Computations for Rotated Magnets
}
\label{sec:computations_of_dipoles}
The optimization of magnet rotations requires many calculations of the magnetic fields of the turned shim magnets, so fast procedures to calculate the magnetic field are needed. The rotations of the $k$ shim magnets are stored as a vector of magnet rotations $\boldsymbol{\alpha} = (\alpha_1,...,\alpha_k)$. Initially, the orientation of all shim magnets is set along the direction of the main field of the magnet, that is along the $\mathbf{z}$ axis of the laboratory frame (Figure~\ref{fig:magnet_with_shim_inserts}) and $\boldsymbol{\alpha}_0=\mathbf{0}$. The magnets are modeled as uniformly magnetized cubes~\cite{ortnerMagpylibFreePython2020} with a given remanence magnetization $B_{rem}$, position and orientation (inset in Figure~\ref{fig:magnet_with_shim_inserts}). To reduce computational time, the magnetic fields are only computed at the location of the acquired field map coordinates. The shim magnets can only rotate in the $yz$ plane and the far-field is assumed ($\vert\mathbf{r}|>>a$ where $a$ is the size of the shim magnet) in the DSV (Figure~\ref{fig:magnet_with_shim_inserts}), that allows using the dipolar approximation. The field of rotated shim magnet can then be calculated as a linear combination of two orthogonal magnetic dipoles ($B_z$, $B_y$). Consider a magnetic dipole moment $\mathbf{m}$ that induces a magnetic field at point $\mathbf{r}$:

\begin{equation}
\label{eq_B}
    \mathbf{B}(\mathbf{r},\mathbf{m})=\frac{\mu_0}{4\pi}\left(\frac{3\mathbf{r}(\mathbf{m}\cdot\mathbf{r})}{\vert r\vert^5}-\frac{\mathbf{m}}{\vert r \vert^3}\right)=\begin{bmatrix}B_y(\mathbf{r},\mathbf{m})\\B_z(\mathbf{r},\mathbf{m})\end{bmatrix}
\end{equation}

Since $\mathbf{B}$ is linear with the dipole moment $\mathbf{m}$, the field generated by a dipole with arbitrary orientation $\mathbf{m}'=(m_z\cos \alpha, m_y\sin \alpha)^T$ can be expressed as a linear combination of precomputed basis fields corresponding to orthogonal dipole orientations: 
\begin{equation}
\label{eq:rotated_dipoles}
    B_z(\mathbf{r},\mathbf{m}') = B_z(\mathbf{r},\mathbf{m})\cos\alpha  -  B_y(\mathbf{r},\mathbf{m})\sin\alpha
\end{equation}
As can be seen in Eq.~\ref{eq:rotated_dipoles}, $B_z$ and $B_y$ can be precomputed once for $\mathbf{m}=\mathbf{0}$, and that is sufficient to rapidly compute $B_z$ for an $\mathbf{m}'$ rotated by any $\alpha$. This allows for a quick evaluation of the field during optimization by avoiding repeated rendering of the field of rotated shim magnets.\\
\par

\subsubsection{Cost Functions}
\label{sec:optimizations_cost_functions}
The vector of magnet rotations $\boldsymbol{\alpha}\in\mathbb{R}^k$ defines the shim field of $k$ shim magnets at $N$ coordinate points: $\mathbf{B}_0(\mathbb{\alpha}) = \mathbf{B}_0 + \boldsymbol{\delta}\mathbf{B}_0+\mathbf{B}_{\text{SHIM}}(\mathbb{\alpha})~\in\mathbb{R}^N$. Here $\mathbf{B}_0$ is the nominal field strength of the magnet array (50~mT for OSI$^2$ ONE magnet) at $N$ measured points $r_i,~i=1\dots N$. $\boldsymbol{\delta} \mathbf{B}_0$ is the deviation of the field to be compensated for, and $\mathbf{B}_{\text{SHIM}}(\boldsymbol{\alpha})$ is the shim field that, ideally, compensates $\boldsymbol{\delta} \mathbf{B}_0$. With the optimization procedure, $\boldsymbol{\alpha}$ is changed so that the corresponding cost function is minimized when $\mathbf{B}_{\text{SHIM}}(\boldsymbol{\alpha})$ approaches $-\boldsymbol{\delta}\mathbf{B}_0$. When a gradient descent, a genetic algorithm or a Monte Carlo search is used to optimize the shim field (Section~\ref{sec:optimizations_positions} and \ref{sec:optimizations_rotations}), a cost function is computed for each $\boldsymbol{\alpha}$.\\
\par
Depending on the optimization algorithm used, the following cost functions were considered. A point-to-point field homogeneity $C_{PTP}(\boldsymbol{\alpha})$ was used for the Monte-Carlo optimization and for the final optimization of the field homogeneity:
\begin{equation}
C_{PTP}(\boldsymbol{\alpha})=\frac{\Delta\text{B}_0(\boldsymbol{\alpha})}{\langle \mathbf{B}_0(\boldsymbol{\alpha})\rangle} \in \mathbb{R}
\end{equation} 
where $\Delta\text{B}_0\in\mathbb{R}$ is the point-to-point inhomogeneity of the shimmed field (max - min), and $\langle\mathbf{B}_0(\boldsymbol{\alpha})\rangle$ is the mean value of the shimmed field.
\\
\par
A vector-valued cost function $\mathbf{C}(\boldsymbol{\alpha})$ was used for the gradient descent algorithm to optimize the squared deviation of the field at all measured points:
\begin{equation}
\mathbf{C}(\boldsymbol{\alpha},\mathbf{r})=\left(\frac{\mathbf{B}_0(\boldsymbol{\alpha},\mathbf{r})}{\langle \mathbf{B}_0(\boldsymbol{\alpha})\rangle} -1\right)^2 \in \mathbb{R^N}
\end{equation}

When a genetic algorithm is used to pick best suitable locations together with the optimal rotations of shim magnets, the cost function is the point-to-point deviation of the field:
\begin{equation}
\label{eq:cost_GA}
C_{GA}(\boldsymbol{\alpha})=\Delta\text{B}_0(\boldsymbol{\alpha}) \in \mathbb{R}
\end{equation}

Additionally, a neural network (NN) was trained to predict the shim configuration (DNA for genetic algorithm) from the field map using the least amount of magnets. To train the self-supervised physics constrained model, the following loss function was used: 

\begin{equation}
\label{eq:cost_nn}
\mathcal{L} = \|\mathbf{B}_{\text{pred}} - \mathbf{B}_{\text{target}}\|^2 + \lambda\,(N_{\text{pred}} - N_{\text{target}})^2 \in \mathbb{R}
\end{equation}

where $\mathbf{B}_{\text{target}}$ is the field map of the target DNA, $\mathbf{B}_{\text{pred}}$ is the field map of the predicted DNA, $N_{\text{target}}$ is the number of magnets in the target DNA, $N_{\text{pred}}$ is the number of magnets in the predicted DNA.

\subsubsection{Optimizing Shim Magnet Rotations}
\label{sec:optimizations_rotations}
When the coordinates of shim magnets are known/fixed, the optimal rotations of the shim magnets are found by minimizing the $L_2$ norm of the vector cost function: 
\begin{equation}
M = \min_{\boldsymbol{\alpha}}\sum_{i=0}^N {C_i}(\boldsymbol{\alpha}) =  \min_{\boldsymbol{\alpha}}\sum_{i=0}^N\left(\frac{\text{B}_0^i(\boldsymbol{\alpha})}{\langle \textbf{B}_0(\boldsymbol{\alpha})\rangle} -1\right)^2    
\end{equation} 
where $\text{B}_0^i$ is $i^{th}$ point of the field map. A vector of magnet rotations $\boldsymbol{\alpha}_x$ that minimizes $M$ is obtained with a least-squares solver based on a trust-region reflective-boundaries (trf) and dogleg box-constrained (dogbox) algorithms, as implemented in SciPy~\cite{virtanenSciPy10Fundamental2020}.\\
After that, the vector of magnet rotations $\boldsymbol{\alpha}_x$ is further optimized using Monte Carlo with $C_{PTP}$ as the cost function. If a deeper minimum is found, the least-squares solver is executed again with the found $\boldsymbol{\alpha}_x$ as an initial guess. When a genetic algorithm is used to optimize the positions and rotations of shim magnets, it is used foremost as it outputs only discrete magnet rotations (4 angles are used between 0 and $2\pi$, cf. Figure~\ref{fig:position_optimization}~d).\\

\par
\subsubsection{Optimizing Shim Magnet Positions}
\label{sec:optimizations_positions}
The shim magnets have to be placed at the most effective positions. The maximum number of used shim magnets is defined before the optimization. The following strategies were implemented and evaluated to distribute the shim magnets by their positions:

    \newpage
    
\begin{figure}[!ht]
    \centering
    \includegraphics[width=
    0.96\linewidth]{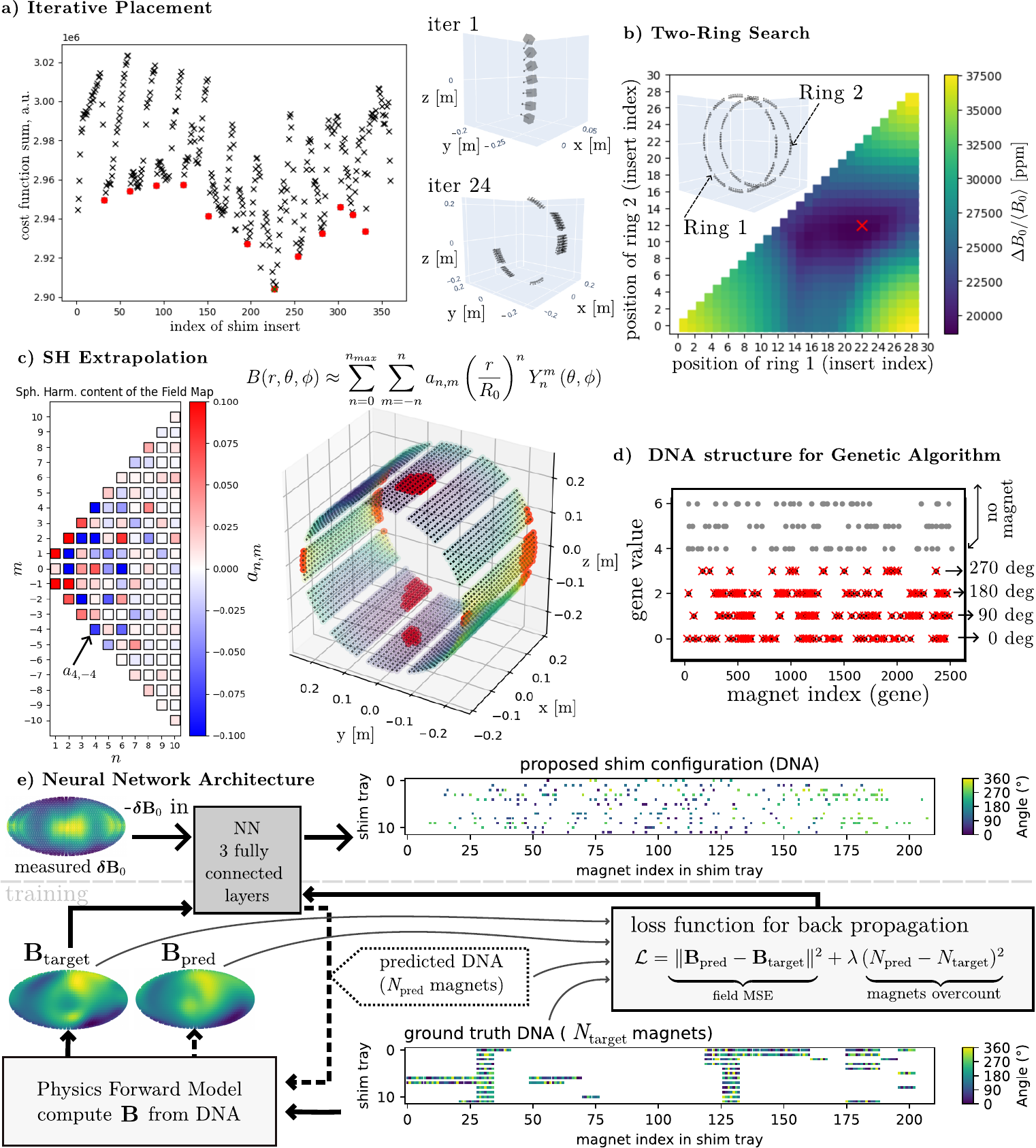}
    \caption{Exemplified results of the different optimization strategies and the functionality implemented in Shimmer. a): iterative placements of magnets or shim inserts. The best 12 positions of the shim magnets are indicated (red dot), that yield the lowest value of the cost function, b): two-ring search, c): Extrapolation of the measured field to the positions of shim magnets using SH, d): best population of the genetic algorithm. Lowest genes correspond to magnet rotations, higher genes correspond to absent magnets, e) architecture of the physics-constrained, self learning NN-based classifier for DNA that was used to instantly predict the shim configuration from the field map.}
    \label{fig:position_optimization}
\end{figure}

\begin{itemize}
    \item \textbf{Iterative placement} (Figure~\ref{fig:position_optimization}~a,~\ref{fig:algorithms_constellations}~a): a probe magnet or alternatively a shim insert holding 7-magnets is added in the simulation at each available position in the shim trays. Rotation of the magnet (or the magnets in the insert) is optimized, $A=\sum_{i=0}^N {C_i}$ is computed for each position. The position with lowest $A$ is occupied with a magnet (or an insert). Then more magnets (or inserts) are added. Process repeats until magnets/trays run out.
    
    \item \textbf{Two-ring search} (Figure~\ref{fig:position_optimization}~b,~\ref{fig:algorithms_constellations}~b): two rings of shim magnets are created at each possible position along the magnet bore. Rotations of magnets in the rings are optimized. Field homogeneity is computed for each pair of shim rings. A pair of rings with best field homogeneity is selected.
    
    \item \textbf{Spherical harmonic extrapolation} (Figure~\ref{fig:position_optimization}~c,~\ref{fig:algorithms_constellations}~c): the measured field map is decomposed into a set of scaled spherical harmonics as described in Figure~\ref{fig:position_optimization}~c) and in Section~\ref{SI_SH}. The obtained coefficients are used for extrapolating the measured field values to the positions of the shim magnets. Shim magnets are added to the positions with most extreme extrapolated field values.
    
    \item \textbf{Genetic algorithm} (Figure~\ref{fig:position_optimization}~d,~\ref{fig:algorithms_constellations}~d): an approach similar to~\cite{wenzelB0ShimmingMethodologyAffordable2021} is used. Shim magnets are created at all available places in the trays (initially 2520 magnets). Both rotations and positions are optimized at the same time. A discrete set of rotations is considered. A gene is introduced to each magnet, that encodes the magnet rotation. If the gene value exceeds a certain value, the magnet is considered as absent. Each shim configuration is represented by a set of genes (an individual). An individual DNA has 2520 genes, corresponding to the state of each available shim magnet. A population of 30000 individuals is evolving for reaching the field with best homogeneity for 100-300 generations, with 85\% crossover, 20\% mutation and 8\% probability of gene mutation. Eq.\ref{eq:cost_GA} is used as the criterion for fitness. The fittest individual corresponds to the optimal magnet placement and rough rotations. The genetic algorithm was implemented using the Deap\cite{shenB1CorrectedBreast2025} package for Python.
    
    \item \textbf{Neural network} (Figure~\ref{fig:position_optimization}~e): a physics-constrained classifier neural network (NN) was trained to predict a DNA used in the genetic algorithm by the measured field distortion $\boldsymbol{\delta}\mathbf{B}_0$. The NN was implemented using PyTorch~\cite{pytorch}, it has 3 hidden layers (length 512) and it was trained using unsupervised physics-controlled learning (no labeled dataset needed). The training data was clustered pseudo-random shim configurations (ground truth DNA), from each test DNA containing $N_{\text{target}}$ magnets, a field map $\mathbf{B}_{\text{target}}$ was computed using the physics forward model (\texttt{OSII\_magnet.compute\_shim\_field\_from\_dna()}). Then the NN was trained to predict the corresponding DNA. From the predicted DNA containing $N_{\text{pred}}$ magnets, the corresponding $\mathbf{B}_{\text{pred}}$ was computed using the same physics forward model. The mean square deviation of the predicted field from the field of the target DNA, together with the mismatch in the predicted number of magnets in the DNA were used as the loss function (Eq.~\ref{eq:cost_nn} and Figure~\ref{fig:position_optimization}~e). The NN was trained for 1000 epochs, that required 12 hours. The training was performed on a workstation equipped with one NVIDIA GeForce GTX 1080 GPU (8~GiB GDDR5X, 2560 CUDA cores, 1733~MHz clock, 320~GB/s memory bandwidth), and one Intel Core i7‑8700K CPU (6 cores, 12 threads, 3.70~GHz clock, 12~MiB L3 cache). During the training a Gaussian noise (2\% RMS) was added to the target field maps to increase the model robustness. Learning rate was 5e-5. The NN gets a field map on its input and predicts a DNA that has information on shim magnet rotations (with a step of 30$^{\circ}$) and their locations (in the shim trays of). Then, gradient descent and Monte Carlo algorithms are used for fine-tuning of the rotations (Figure~\ref{fig_SI_homo_vs_nmagnets}, red cross). The NN minimizes both error of the predicted field and the number of used magnets in the resulting DNA, therefore, to get solutions with larger amount of magnets the NN has to be further trained. Number of magnets used in the solution can be decreased on demand.

\end{itemize}

After positions $\boldsymbol{\rho}_x$ and rotations $\boldsymbol{\alpha}_x$ of the shim magnets are optimized, the expected field map is computed and the 3D printable files for the magnet holders are generated.

\subsection{Measurement Setup}

\begin{figure}
    \centering
    \includegraphics[width=1\linewidth]{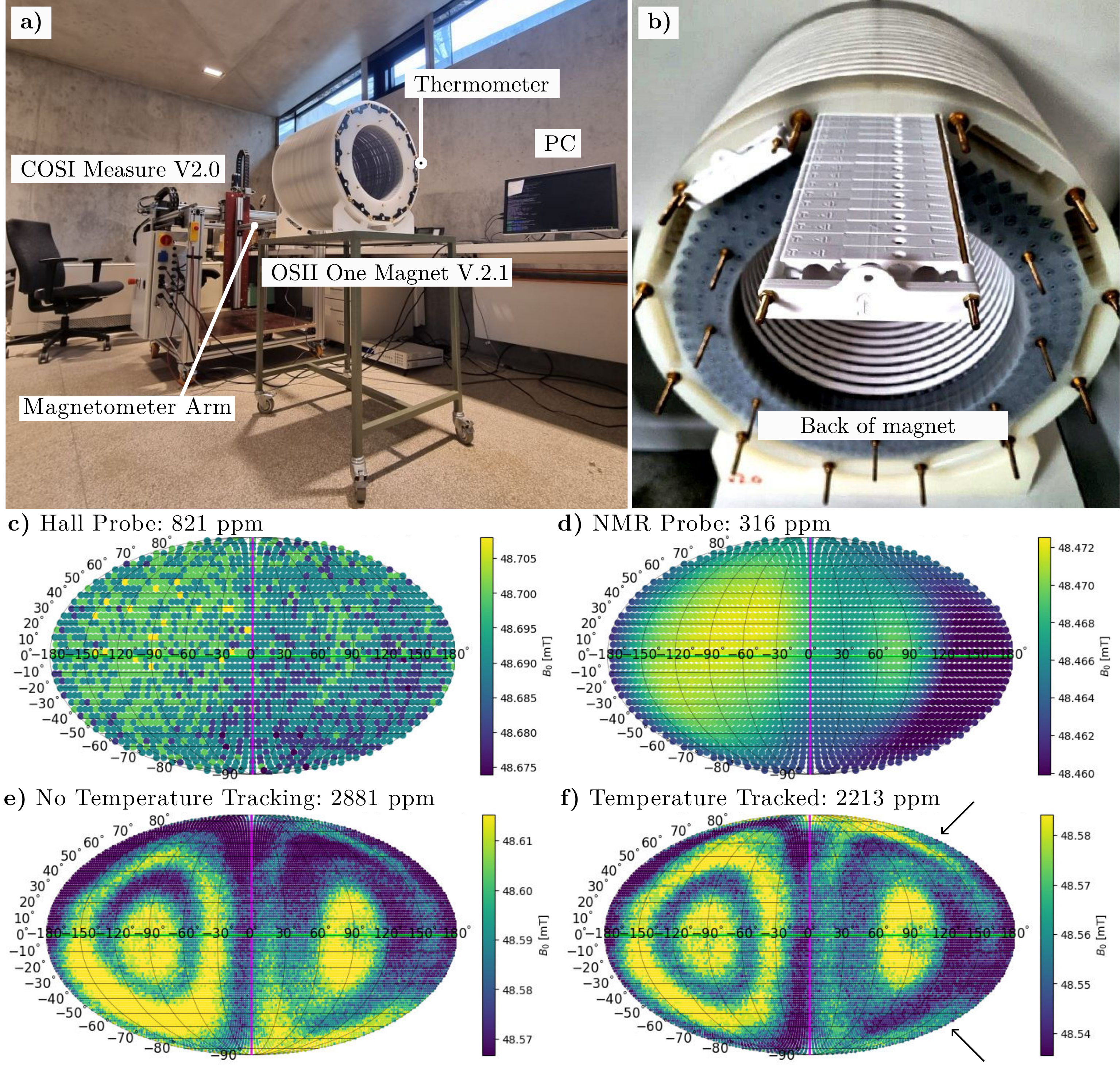}
    \caption{Magnetic field mapping measurement setup. a): COSI Measure equipped with a magnetometer placed wtthin the OSI$^2$ ONE low-field MRI magnet v2.0. b): Assembled shim trays in the magnet. c,d): $\mathbf{B}_0$ field maps recorded on a 100~mm DSV using two different field probes. e,f): Temperature tracking during an overnight field mapping, 16324 points on a 200~mm DSV. left: field map without temperature tracking, right: probe readings corrected by the tracked temperature drift. Visible field changes after applying temperature compensation shown in arrows.}
    \label{fig:measurement_setup}
\end{figure}

\subsubsection{Magnetic Field Mapping}
Mapping of the magnetic field was performed using the positioning system COSI Measure v2.1~\cite{hanOpenSource3D2017,cosi_gitlab}, equipped with a 3-axis Hall-effect magnetometer (Model 460, LakeShore Cryotronics, Westerville) and a NMR magnetometer (Model PT2026 Precision Teslameter, Metrolab Technology SA).  An experimental setup is shown in Figure~\ref{fig:measurement_setup}~a). The Hall-effect sensor is used for the field mapping of the most inhomogeneous magnets e.g. for the first shimming iteration. This allows to map bigger DSVs since the NMR magnetometer requires a field homogeneity of around 1000~ppm at the NMR probe tip (Figure~\ref{fig:measurement_setup}~c), Figure~\ref{fig_SI_SHIMMING_PROGRESS_MAGNET_V1}), which is difficult to obtain on the surface of the DSV in an inhomogeneous magnet. The quantization noise of the Hall-effect magnetometer used is however in the range of 1000~ppm, so after 1-2 iterations the NMR magnetometer was applied, which has a higher measurement sensitivity (Figure~\ref{fig:measurement_setup}~c).

\par
The magnetic field inside a closed volume, that does not contain any magnetic dipoles, can be computed from the values of the magnetic field on the surface of that volume using scaled spherical harmonics~\cite{wenzelB0ShimmingMethodologyAffordable2021}. Furthermore, since the permanent magnets are placed outside of the probed sphere, the inhomogeneity of the field is the highest at the surface of the DSV and only decreases towards the center of the sphere. That implies, that it is sufficient to map the field only at the surface of the DSV, which reduces the number of mapped points and the overall measurement time. 
The trajectory of the magnetometer probe follows a spiral path on the surface of the sphere. The points on the surface are spaced equidistantly.
\par
For the measurement procedure typically 2.5~s per measurement point are set, to avoid errors due to averaging times or vibrations of the probe. The selected number of points on the spherical surface should be dense enough to capture inhomogeneities, but not too dense to keep measurement times low and avoid strong magnetization drifts of the permanent magnets due to temperature changes while mapping (Figure Figure~\ref{fig:measurement_setup}~d). Here the spacing of mapped coordinates on the surface was 3.2~mm, which resulted in a total of 16324 measurement points and 8.23h to measure a surface of a 200~mm DSV. In order to compensate for the drifting $B_{rem}$ during the measurement of the field map, the temperature of the magnet was monitored with the Luxtron M-900 fiber optic fluorescent temperature probe positioned at the outer surface of the magnet. 
In Figure~\ref{fig:measurement_setup}~d) a significant drift in $\mathbf{B}_0$ is visible due to the change in temperature by $\Delta T=0.90\pm0.01~$K. A temperature drift compensation was added to later measurements to account for these errors, which can be in the range of 600~ppm, while a repeated measurement of the same field map with the same (NMR) probe results in an error within 100~ppm.

\begin{figure}[h]
\centerline{\includegraphics[width=\textwidth]{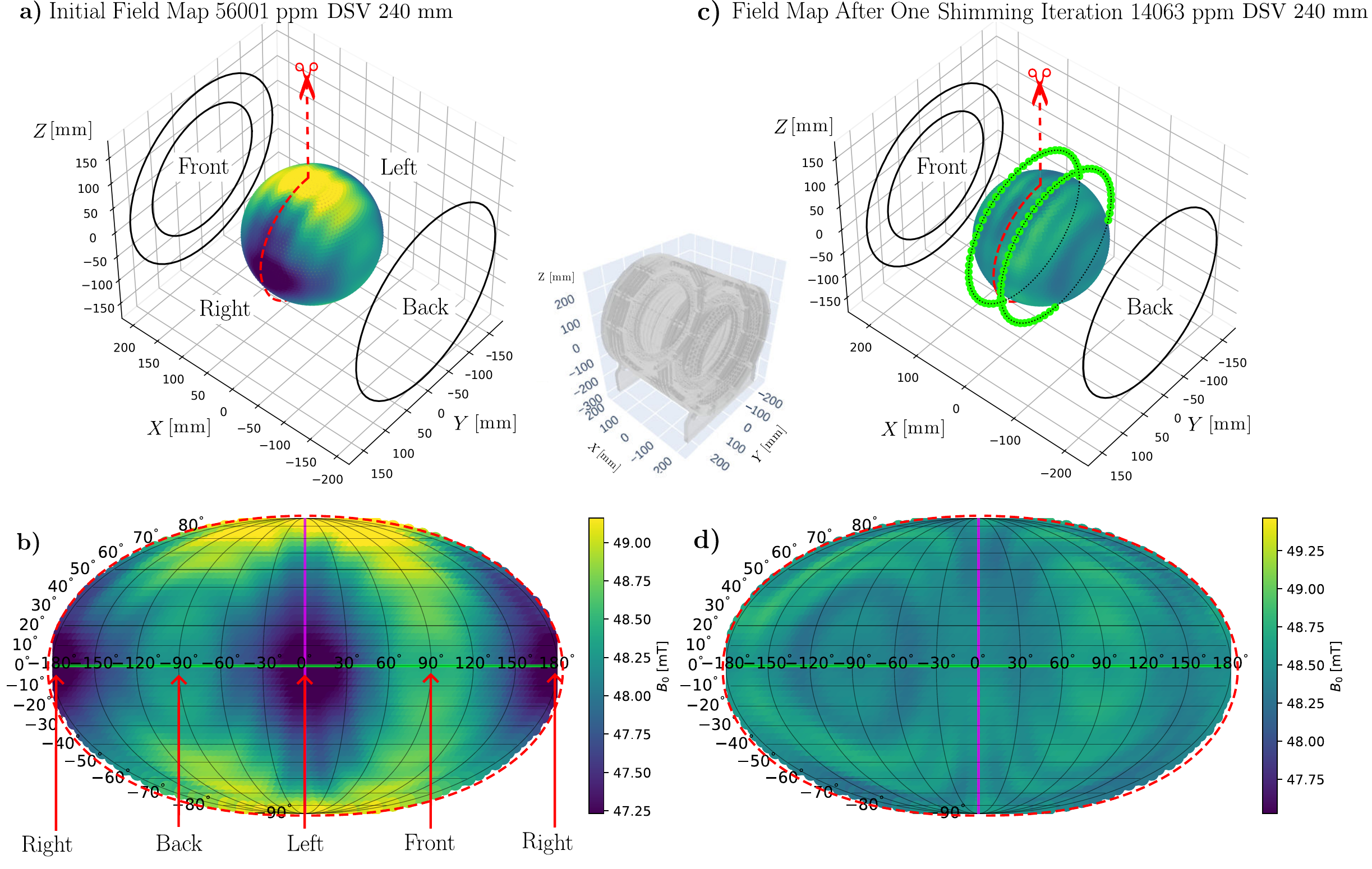}}
\caption{a): Initial map of the $\mathbf{B}_0$ field recorded on a surface of a 240~mm DSV inside the unshimmed magnet b): $2D$ Mollweide projection of the initial field map before shimming. The cut is shown in a dashed line. c): field map after one shim iteration with two rings of 9~mm magnets. Shim rings are shown in green. d) Mollweide projection of the shimmed field map shown in the same data range as the unshimmed map.\label{fig_mollweide_cuts}}
\end{figure}

The field map recorded on a surface of a sphere is then mapped onto a plane with a Mollweide projection for a comprehensive representation (Figure~\ref{fig_mollweide_cuts}). 

\par
If the coordinate system of the field mapping device does not match exactly the coordinate system of the magnet, a script in Shimmer can be used to do a proper coordinate transformation bringing the measured data into magnet coordinates for successful shimming optimizations. The coordinate system used for the OSI$^2$ ONE v2.2 magnet is exemplified in Figure~\ref{fig:magnet_with_shim_inserts}. The principal component of the magnetic field in the bore points along the $+z$ axis, the bore extends along the $x$ axis. The rotation of the shim magnets is adjusted in the $yz$  plane. The initial orientation of all shim magnets in the simulation is along the $+z$ axis: $\mathbf{m}_i\uparrow\uparrow \vec{\mathbf{z}}$. During the simulation, each shim magnet is rotated about an axis that passes through its center and is parallel to the magnet``s bore direction ($\mathbf{x}$), so that $B_y$ and $B_z$ field components of each shim magnet change. Only the principal, $B_z$ component is optimized in the simulation. 

\subsection{MR Imaging}
The inhomogeneity of $\mathbf{B}_0$ is directly related to the Larmor frequency and impacts MR image quality. To demonstrate the influence of field inhomogeneity on MR image quality, MR imaging was performed before and after $\mathbf{B}_0$ shimming. MR imaging was performed on an open-source low-field MRI Scanner~\cite{osii_one_scanner_gitlab} equipped with an OSI$^2$ ONE v2.1 magnet. A 150 mm, 20-turn solenoid RF coil~\cite{rf_coil_gitlab} was used and loaded with the 110-mm OSI$^2$ Hello World phantom~\cite{helloworld_phantom}. The pulse sequence was a 3D turbo spin echo (TE/TR = 14/600~ms, ETL = 10, FOV = 150~mm, spatial resolution = 1.5$\times$1.5$\times$15~mm, readout bandwidth = 20~kHz, acquisition time = 63~s), which was executed on the open-source Nexus console~\cite{schoteNexusVersatileConsole2025a}.

\section{Results}\label{results}

\subsection{Optimization Strategies}
\begin{figure}[h]

\centerline{\includegraphics[width=0.9\textwidth]{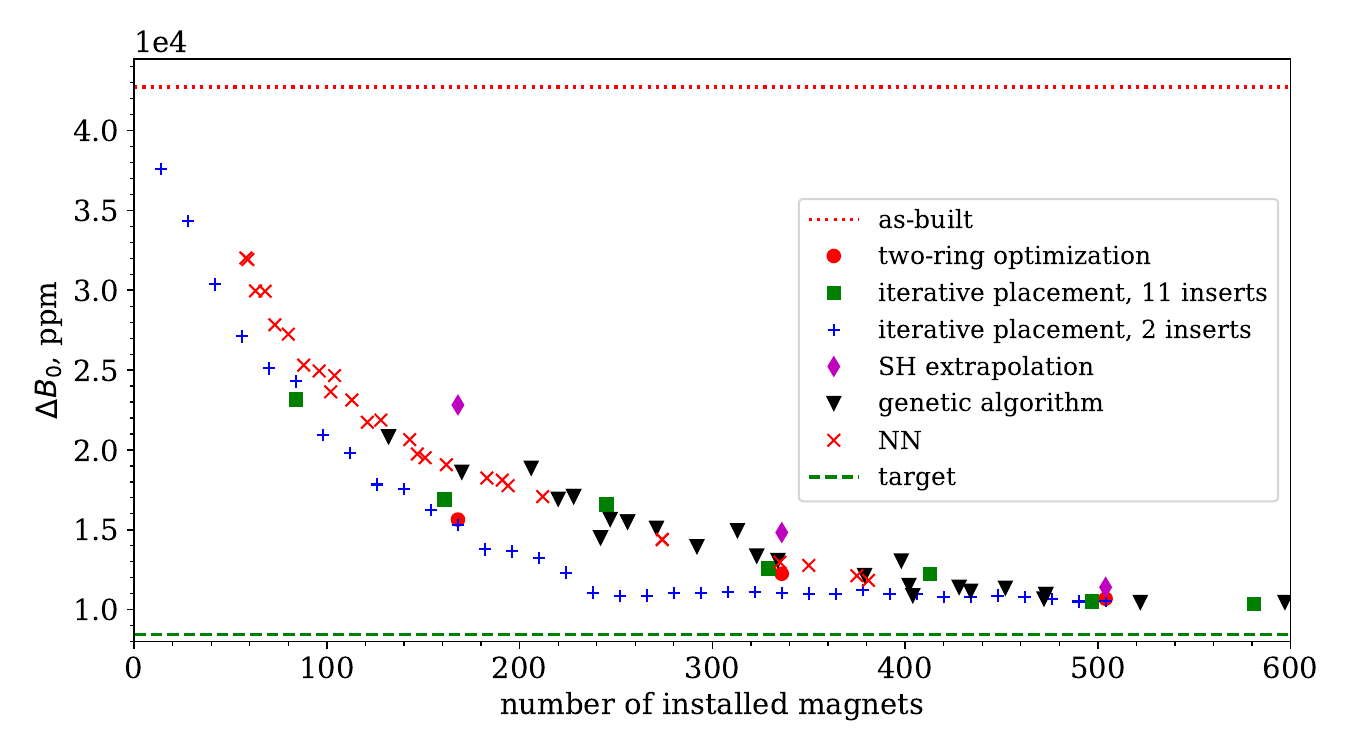}}

\caption{$\mathbf{B}_0$ inhomogeneity (240 mm DSV) in the as-built OSI$^2$ ONE magnet v2.0 as a function of the number of installed shim magnets using different optimization methods (simulation). Initial inhomogeneity of the as-built magnet is indicated as red dotted line (42726~ppm). Target field homogeneity is indicated as a green dashed line (8460~ppm, as simulated for the ideally assembled magnet, see Figure~\ref{fig:opt_vs_dsv} and Supplementary Section~\ref{SI_expected_homos}). Iterative placement of single magnets by searching best magnet position is shown in blue crosses (2 7--magnet inserts in one optimization step) and in black crosses (11 7--magnet inserts in one optimization step). Placement of magnets in 12-insert rings (84 magnets per ring) is shown in red dots. Placement of magnets by using genetic algorithm is shown in black triangles, placement of magnets with the neural network (NN) is shown in red crosses, and the placement of magnets at the most affected positions computed with the SH extrapolation of the distortions in the field map is shown in magenta diamonds.\label{fig_SI_homo_vs_nmagnets}}
\end{figure}

\begin{figure}[ht!]
    \centering
    \includegraphics[width=1\linewidth]{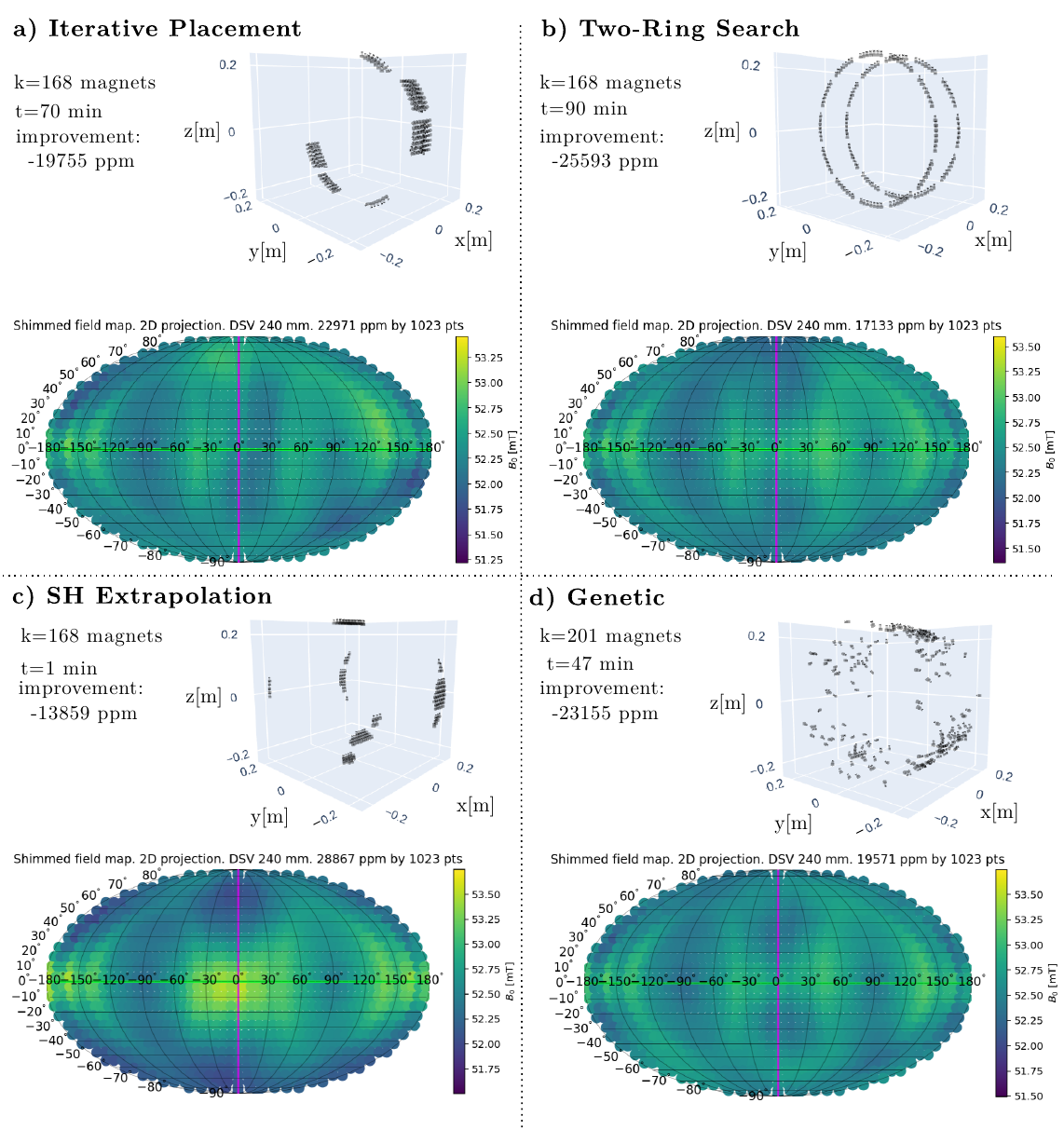}
    \caption{Comparison of the performance of the four optimization strategies for $\mathbf{B}_0$ shimming for one iteration with a defined number of magnets. The initial field map has a field inhomogeneity of 42726~ppm (DSV 240 mm). Time of computation, number of used magnets and improvement of homogeneity is shown for each algorithm. a): iterative placement of shim inserts (7 magnets in one insert), two inserts in one optimization step, b): two-ring search, c): Spherical harmonic extrapolation, d): genetic algorithm. All field maps are shown in the color scale of the original unshimmed map.}
    \label{fig:algorithms_constellations}
\end{figure}

The results of the different optimizations strategies implemented in Shimmer are shown in Figure~\ref{fig_SI_homo_vs_nmagnets} and Figure~\ref{fig:algorithms_constellations}. The as-build field map (DSV 240~mm) of the OSI$^2$ ONE magnet v2.0 was used as the test dataset. 

As the number of magnets ($k\approx 400)$ increases, all optimization strategies are converging to similar $\mathbf{B}_0$ inhomogeneities of around 10000~ppm, which is close to the inhomogeneity of the ideally simulated magnet of 8460~ppm (240~mm DSV, Figure~\ref{fig_SI_expected_homos}), rendering all optimization approaches feasible. The fastest convergence is achieved with the iterative placement of single inserts (two 7-magnet inserts in one optimization step), where the least amount of magnets (k=250) yields similar results versus all other approaches (k$>400$). This is important, since next to the savings in magnet cost and weight, more free space is available within the shim trays for the consecutive shim iteration. The iterative placement of 11 fully populated shim inserts (green squares in Figure~\ref{fig_SI_homo_vs_nmagnets}) and the two-ring approach (red circles), also perform reasonably well with the least number of magnets used. To better compare this, at a fixed number of shim magnets that can be used for one iteration (168), the two-ring optimization outperforms the iterative placement of shim inserts, the SH and the genetic algorithm, especially at earlier shimming iterations (Figure ~\ref{fig:algorithms_constellations}). The fastest optimization approaches to find an improved shim configuration is the SH extrapolation and the NN (not shown). For 168 magnets, the SH computations needed less than 1~min, in comparison to 47-90~min for the other optimization strategies (Figure ~\ref{fig:algorithms_constellations}~c). Similarly, the neural-network based classifier for the DNA provided a very quick computation ($t<5$~s) of the DNA from any input field, once the NN was trained. Even though the SH extrapolation is a less effective method regarding the number of used shim magnets, it can be a fast approach to investigate if and how good a particular magnet design can be shimmed in a given shim magnet configuration. The genetic algorithm comes close in its performance to the two-ring search and can be in particular useful at later shimming iterations to fully populate all shim tray locations and further approach the ideal shimming target. \\

\subsection{Validation Measurements}
\label{sec:simulation_performance}

\begin{figure}[h]
\centerline{\includegraphics[width=\textwidth]{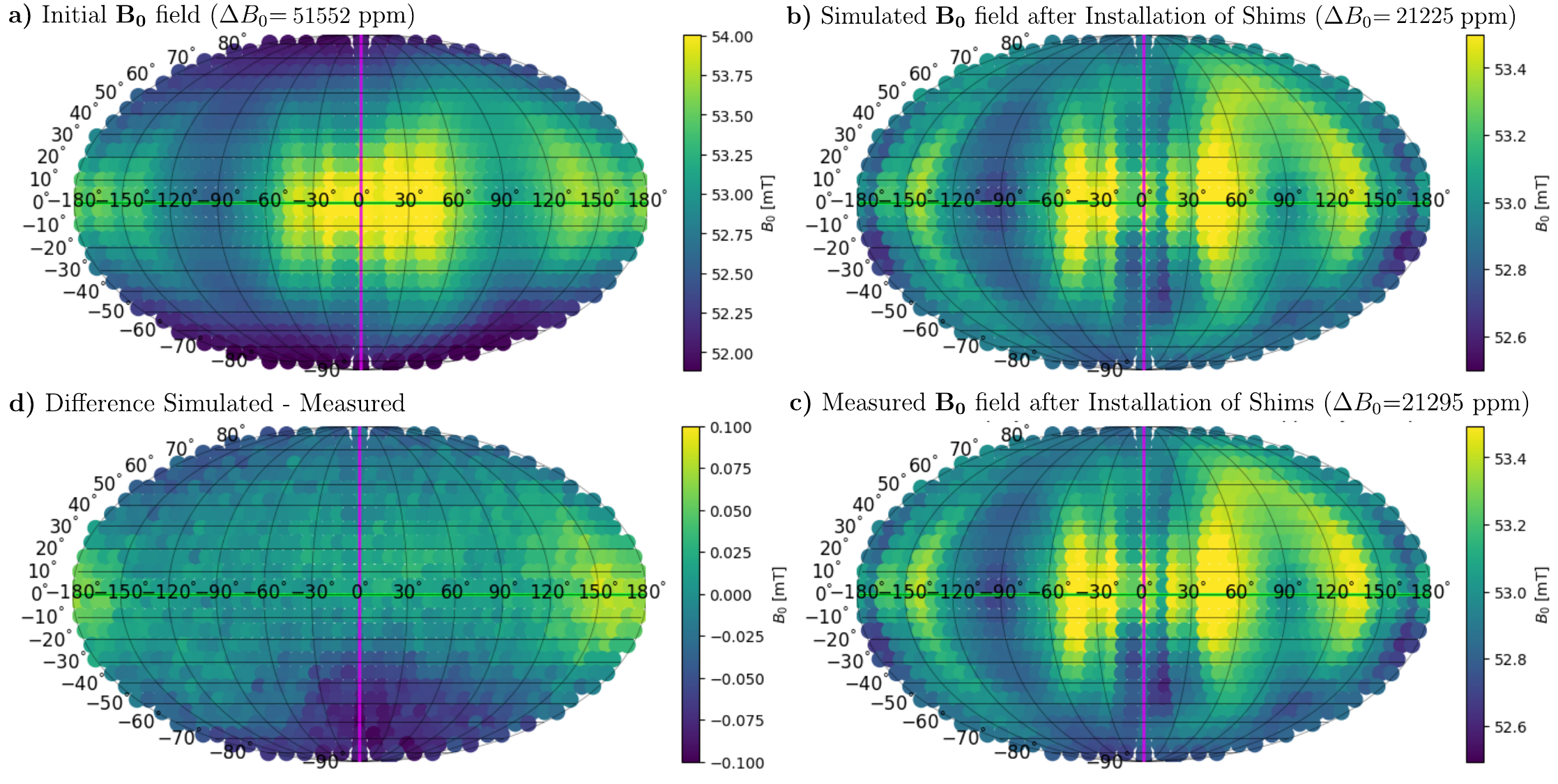}}
\caption{Comparison between the simulated and measured shimmed $\mathbf{B}_0$ field after one shim iteration for the OSI² ONE v2.0 magnet. Field maps were recorded on a 240 mm DSV with a Hall magnetometer. a) Initial field map of an as-built magnet, field homogeneity 51552~ppm b): simulated shimmed field with two shim rings containing 168 8~mm NdFeB magnets grade N45. Expected field homogeneity 21225~ppm c): measured field map after the installation of the shim magnets. Measured field homogeneity 21295~ppm, d) difference between the simulation and the measurement.  \label{fig3}}
\end{figure}

After installing the shim magnets in the shim inserts, the shim inserts in the shim trays and the shim trays in the MR magnet, the temperature-compensated measured shimmed field maps were compared to the simulated fields using Shimmer. The measured results are displayed in Figure~\ref{fig3} and show a good agreement with the simulations. While simulations implementing one shim iteration predicted a reduction of $\mathbf{B}_0$ field inhomogeneity by a factor of 2.43 of the initial value (from 51552~ppm to 21225~ppm), the measurements showd field inhomogeneity of 21295~ppm which is only 70 ppm higher than the expected value (Figure~\ref{fig3}~b,c). The difference map (Figure~\ref{fig3}~d) between the simulation and measurement indicates absolute peak-to-peak deviation of $0.187$~mT, which translates to 3522~ppm when computed with respect to the mean measured field.

\begin{figure}[!ht]
\centerline{\includegraphics[width=1\textwidth]{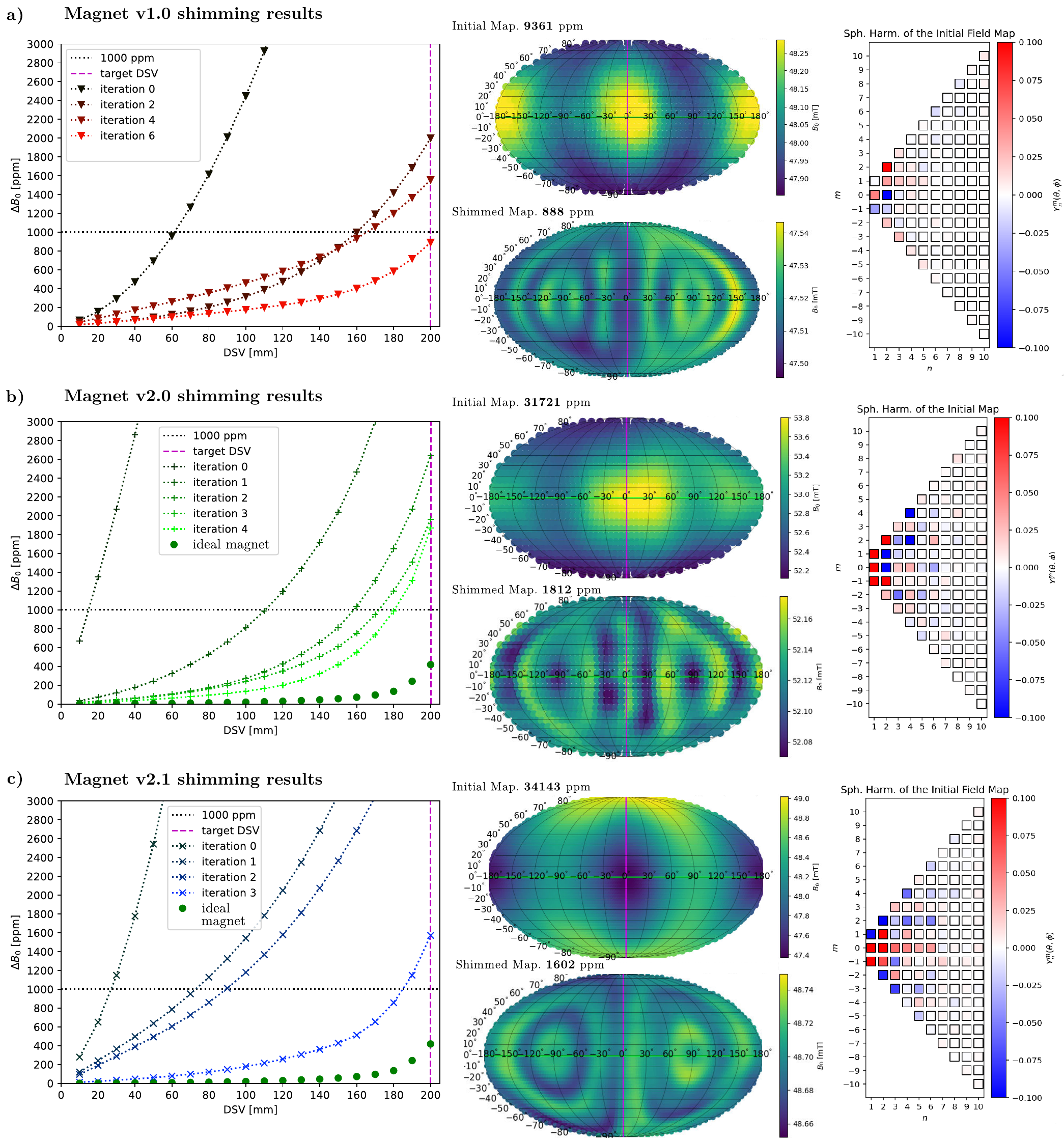}}
\caption{$\mathbf{B}_0$ shimming results using Shimmer on three different MR magnets: a) OSI² ONE v1.0, b) OSI² ONE v2.0 and c) OSI² ONE v2.1. left: $\mathbf{B}_0$ homogeneity as a function of the probed spherical volume for different shimming iterations. Middle: $\mathbf{B}_0$ field maps (DSV 200~mm) before and after the passive shimming. Right: spherical harmonics present in the field map before shimming (as-built). Field maps, spherical harmonics coefficients and shim magnet placement for all magnets and each iteration can be found in the supplementary figures Figure~\ref{fig_SI_SHIMMING_PROGRESS_MAGNET_V1} (v1.0), Figure~\ref{fig_SI_SHIMMING_PROGRESS_MAGNET_V2p0} (v2.0) and Figure~\ref{fig_SI_SHIMMING_PROGRESS_MAGNET_V2p1} (v2.1) \label{fig:homo_vs_dsv}}
\end{figure}

To investigate the cause for potential deviations, simulations were used to study how strongly deviations of shim magnet parameters affect the homogeneity of a shim field. For this purpose 2520 8-mm N45 magnets occupying all shim trays were used and simulated with varying shim magnet rotation, shim magnet positions and shim magnet remanence on a DSV 240~mm for the OSI$^2$ ONE v.2.1 magnet (Section~\ref{SI_shim_magnet_uncertainties}). The shim field uncertainty stays within 1000~ppm if shim magnet rotation is $\Delta \phi\leq$2$^\circ$, shim magnet position is $\Delta x\leq$3~mm and shim magnet remanence is $\Delta B_{rem}\leq$3~\%. In comparison, similar calculations of variations in the main magnet array yield much lower tolerances: $\Delta \phi_{PM}\leq$0.3$^\circ$, $\Delta x_{PM}\leq$0.3~mm, $\Delta B_{rem PM}\leq$0.6~\%, Section~\ref{SI_error_sources}.

\subsection{$\mathbf{B}_0$ Shimming of Three MR Magnets}
After successful validation measurements, the $\mathbf{B}_0$ shimming procedure was applied to three different in-house built OSI$^2$ ONE magnets (v1.0, v2.0 and v2.1). The results are illustrated in Figure~\ref{fig:homo_vs_dsv} with the detailed results for each shim iteration in Figures \ref{fig_SI_SHIMMING_PROGRESS_MAGNET_V1},\ref{fig_SI_SHIMMING_PROGRESS_MAGNET_V2p0},\ref{fig_SI_SHIMMING_PROGRESS_MAGNET_V2p1}. 
The initial field inhomogeneities of 9361~ppm for v1.0, 31721~ppm for v2.0 and 34143~ppm for v2.1 within a 200~mm DSV were improved to 888~ppm, 1812~ppm and 1602~ppm respectively within 3-6 iterations. The resulting DSV with field inhomogeneities below 1000~ppm was 200~mm for magnet v1.0, 180~mm DSV for magnet v2.0 and 185~mm for magnet v2.1. All three magnets showed distinct initial field patterns, which are visible in the initial field maps and their corresponding SH decomposition. Nevertheless, in all these configurations, Shimmer performed reasonably well to homogenize $\mathbf{B}_0$. Similar to the convergence observed in Figure~\ref{fig_SI_homo_vs_nmagnets}, the initial shimming iterations for all magnets yield the strongest changes in field homogeneity (Figure~\ref{fig:homo_vs_dsv}), while the improvements from the last iterations, when $\mathbf{B}_0$ field homogeneity reaches values below 2000~ppm, is less pronounced. It can be observed that within the first iterations the lower spherical harmonic orders are effectively reduced, leading to the observed improvements in field homogeneity (Figure~\ref{fig_SI_SHIMMING_PROGRESS_MAGNET_V2p0}). Ultimately, higher order terms remain, which are also visible in the concentric ring patterns visible in the magnetic field map of the last shim iteration (Figure~\ref{fig:homo_vs_dsv}). These ring patterns correlate with the physical magnetic ring location of each magnet design (Figure~\ref{fig_SI_expected_homos}).

\subsection{Optimization Results vs. DSV}

\begin{figure}[!ht]
\centerline{\includegraphics[width=1\textwidth]{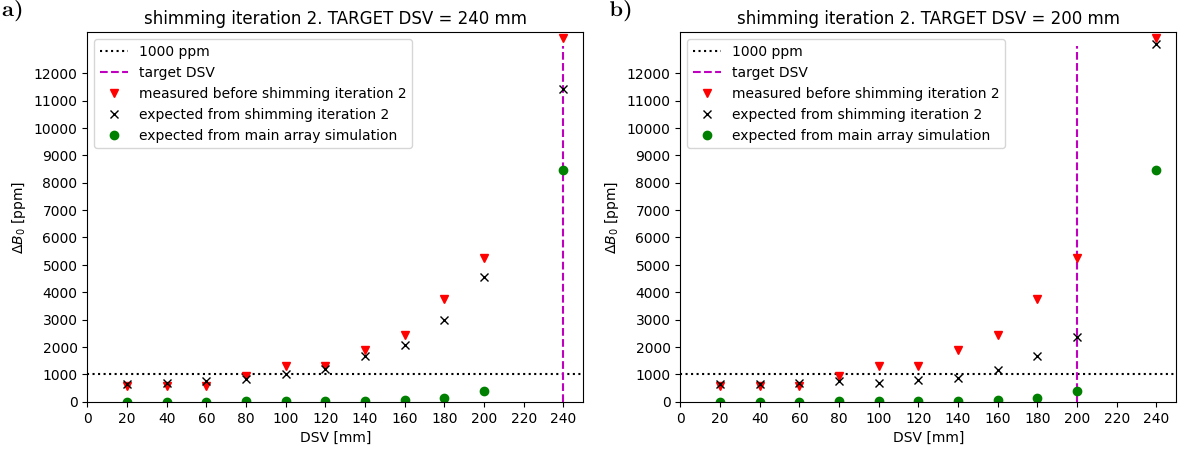}}
\caption{Changing the target DSV for improving the overall field homogeneity for OSI$^2$ ONE magnet v2.0, passive shim iteration 2. a): target DSV 240~mm, b): target DSV 200~mm. Both optimizations use the same number of magnets (168 8~mm N45 magnets).\label{fig:opt_vs_dsv}}
\end{figure}

The optimization of the field is performed on a fixed DSV and as already seen in Figure~\ref{fig:homo_vs_dsv}, the optimization effects are less pronounced at later shim iterations, when the intrinsic effects of the magnet design become dominant and additional shim magnets of limited size cannot compensate them anymore effectively. To demonstrate the effect of the choice of DSV on the shimming performance, Figure~\ref{fig:opt_vs_dsv} shows a comparison of different DSVs used for shimming iteration 2 of magnet v2.0. Both plots correspond to the same number of shim magnets used in the optimization (168 8 mm magnets grade N45). Iterative placement of 2 inserts was used to find the best magnet positions. When a larger DSV is chosen for the optimization (Figure~\ref{fig:opt_vs_dsv}~a), the field remains inhomogeneous at smaller DSVs, because the shim magnets are not strong enough to compensate the field distortions closer to the magnet rings. However, when the same optimization algorithm is applied to a smaller DSV (Figure~\ref{fig:opt_vs_dsv}~b), the overall field homogeneity improves. This happens because at a smaller DSV the characteristic field distortions from the main magnet rings (Section~\ref{SI_expected_homos}) become comparable to the shim fields, allowing the smaller shim magnets to compensate the distortions caused by the larger magnets of the main array.

\subsection{MR Imaging Performance}
\begin{figure}[h]
\centerline{\includegraphics[width=0.86\textwidth]{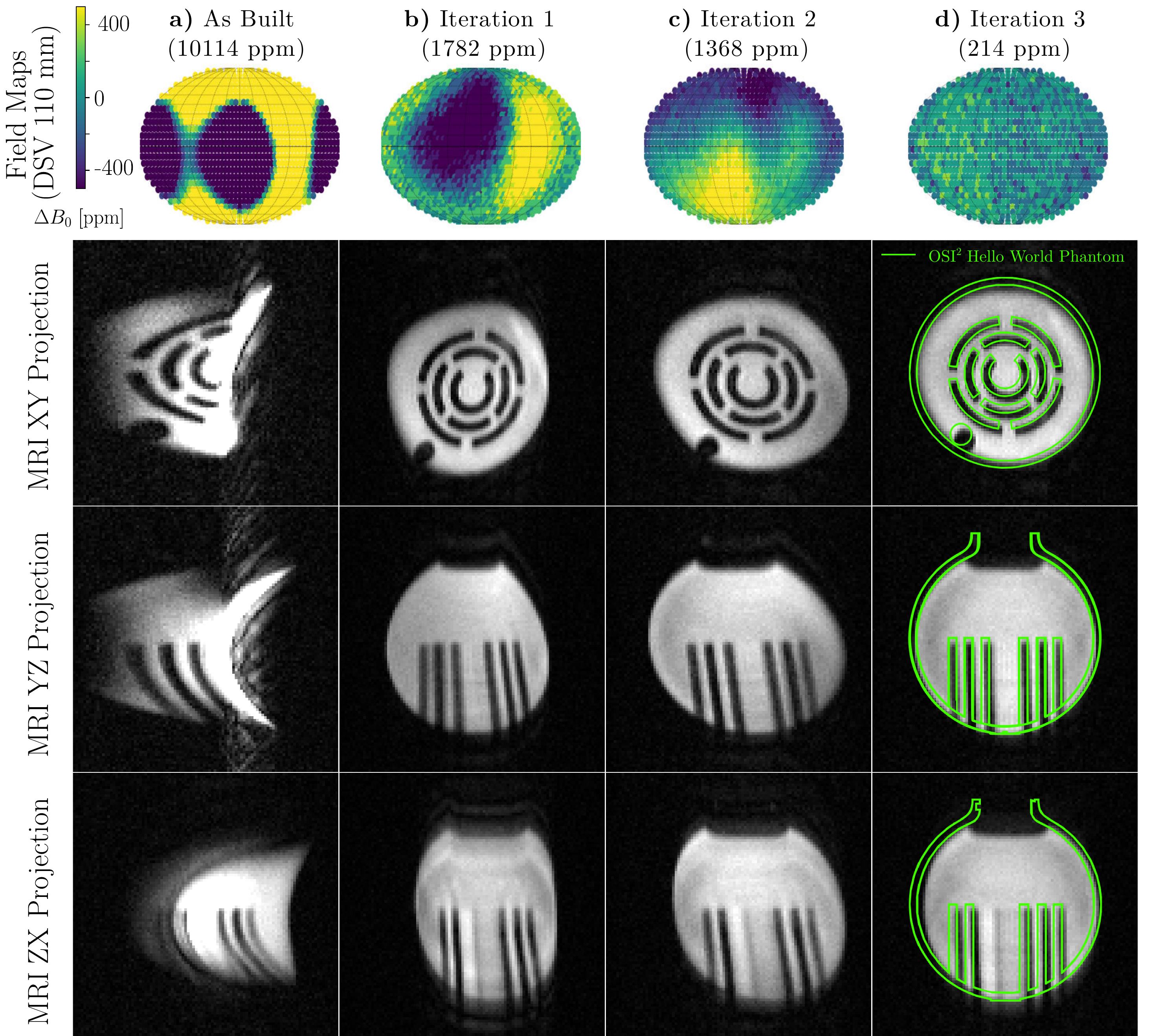}}
\caption{$\mathbf{B}_0$ field maps (DSV 110 mm) and MR images at three successive shimming iterations of the OSI$^2$ ONE v2.1 50~mT low-field MRI magnet. As a test object 110-mm the spherical OSI$^2$ Hello World phantom~\cite{helloworld_phantom} was positioned at the isocenter of the MR magnet.\label{fig4}}
\end{figure}

The MR imaging results before and after shimming are displayed in Figure~\ref{fig4}. For the initially magnet build prior to shimming heavy distortions, image artifacts and incomplete RF excitation are visible that relate to the strong inhomogeneities of 10114~ppm that are present in the target DSV of 110~mm. At these inhomogeneities MR imaging is challenging. After only one iteration and an improvement of an order of magnitude to $\Delta\text{B}_0=1782$~ppm (Figure~\ref{fig4}~b) RF excitation covers the full phantom and the structure becomes visible even though some remaining distortions remain. Similarly at this stage the FID signals and SNR increase. At the last iteration, only minor distortions are visible and the object compares well to the theoretical dimensions of the phantom (Figure~\ref{fig4}~d).

\section{Discussion}\label{discussion}
Shimmer was successfully applied to improve the $\mathbf{B}_0$ field homogeneities of three different magnets to below 2000~ppm leading to improvements by factors 10-21 from the unshimmed initial fields. Other groups were successfully reproducing these results by using the open-source base to shim their magnets and even perform in-vivo imaging~\cite{pfitzerFENCEFlexibleElectric2026}.\\
\par
Shimmer enables the use of different optimization strategies for placing the shim magnets when simulating the optimal shim fields. For the first shimming iterations, the two-ring search or the iterative placement of shim inserts appear to be the most effective placement strategies for the as-built magnets, reducing the initial inhomogeneity substantially while using the available shim magnet space effectively. At later shimming iterations individual magnet placement or genetic evolution may deliver better performance in terms of used magnets. After initial training, the DNA generated by the trained NN can be used as population seeds for the genetic algorithm, yielding similar shimming performance with a substantial increase in computational performance, which further helps to e.g. optimize the amount of magnets used in the shim configuration. The approach with spherical harmonic decomposition allows for very fast computations that can be useful e.g. to estimate if a certain shim tray or magnet design can be used effectively to homogenize the field, however its overall performance is worse compared to other methods. In different applications, this quick-computation approach might still be useful e.g. to compute the positions of ferromagnetic sheets inside the magnet bore to further improve the field homogeneity~\cite{zhaoNovelPassiveShimming2025}.\\

\par
Shimmer is based upon the \texttt{OSII\_magnet.py} class, that can be easily adapted for testing other optimization approaches. The source code of Shimmer comes with a mock up script that summarizes all available optimization algorithms and allows for switching between different implementations.
Shimmer can also simulate the fields of the main magnet array, which allows for several useful applications. One such application is the ability to detect incorrectly inserted magnets in both the main magnet array and the shim inserts. It furthermore allows to estimate the theoretical homogeneity intrinsic to the magnet design, which is useful to estimate the performance of any shimming strategy or apply the code to improvements of the main magnet design.\\ 

\par
Zanovello et al. showed~\cite{zanovelloVeryLowFieldMRIScanners2025} that the difference in field inhomogeneity between simulation and construction of the OSI$^2$ ONE v2.0 magnet is largely due to a slight angular dislocation of the permanent magnets due to the torque that emerges between adjacent rings where the magnets are installed. Due to the Halbach arrangement, stronger magnetic forces are expected inside the magnet, while on the outside, where the shim trays are positioned, the magnetic forces largely cancel out indicating that the angular dislocation of the shim magnets due to those forces has a minor effect as compared to the main magnet. Nevertheless, in order to investigate the influence of uncertainties of shim inserts positioning on the $\mathbf{B}_0$ field homogeneity, computations were performed and are shown in the Supplementary Section~\ref{SI_error_sources}. It can be seen (Section~\ref{SI_error_sources},~\ref{SI_shim_magnet_uncertainties}) that for the main magnet array, changes in magnet rotation of more than $0.3^{\circ}$, position errors of more than 0.3~mm and deviations of remanence $B_r$ more than 0.6\% (standard distribution) lead to deviations in $\Delta\text{B}_0>1000$~ppm. Shim magnets, however, being placed further away from the DSV, have much more relaxed criteria on precision as compared to the main magnets (rotation error $2^{\circ}$, position error $3$~mm, $B_{rem}$ error 3\%, Section~\ref{SI_shim_magnet_uncertainties}). Nevertheless, these uncertainties including the measurement uncertainties of the magnetic field probes, explain the challenges at later iterations to improve the $\mathbf{B}_0$ field homogeneity further to values below 1000~ppm. 
The standard deviation for magnet $B_{rem}$ is typically higher than 0.5\% and depends on the manufacturer and also production batch. Individual measurements and sorting of magnets, as performed in~\cite{wenzelB0ShimmingMethodologyAffordable2021}, can help to further reduce this error e.g. by using an open-source magnet test station ~\cite{zanovello_magnet_test_station}. Position and rotational errors are more difficult to correct with simple production techniques. By choosing the location for the shim trays outside Halbach-like magnets, magnetic field contributions from the main magnet are e.g. relatively small, leading to only small or minor forces on the assembled shim magnets that could lead to displacement or rotational deviations. With respect to precise data acquisition, temperature tracking and corrections need to be implemented, as also performed in \cite{galveEllipticalHalbachMagnet2024}. Additionally, while for the first shimming iterations the sensitivity of the Hall-effect magnetometer might be sufficient, later iterations, where the measured field homogeneity reaches values of 1000~ppm, might require more sensitive field probes such as NMR based magnetometers. At this stage however, when the $\mathbf{B}_0$ field is already homogeneous enough for MR imaging applications, $\mathbf{B}_0$ mapping might allow to further optimize the shimmed fields.
\\
\par
The validation measurements in Section~\ref{sec:simulation_performance} show a maximum absolute deviation between simulated and measured field of 0.187~mT or 3522~ppm, which corresponds to an error of 11.6\% with respect to the absolute expected decrease in homogeneity of 30327~ppm in this shim iteration. For earlier shim iterations of an inhomogeneous magnet this mismatch due to the earlier mentioned uncertainties is less critical and the simulated $\mathbf{B}_0$ homogeneity matches the measured values (only 70~ppm difference) very closely. For later shim iterations, however, its impact increases and it gets increasingly more challenging to achieve larger improvements within a single shim iteration.

\par
The generic inserts shown in Figure~\ref{fig_SI_blueprints}~d) are useful to easily change magnets and/or recycle them. This also enables an reoptimization of the shim fields by e.g. recomputing the shimmed magnets used after the last iteration and e.g. removing shim magnets or changing their orientation/size. In the current version a whole shim insert with e.g. 7 populated magnets is inserted. The iterative search of single magnets showed however, that an individual placement of magnets might be more beneficial, especially at later shim iterations. Here an easy modification of the design can be performed to individually place magnets withing the shim insert\\

\par
For all three magnets shimmed, it can be seen that when approaching 1000~ppm, the characteristic ring pattern of the built magnet becomes visible, which is difficult to compensate further using the shim architecture proposed in this work. Here other techniques might help, such as placing patterned ferromagnetic sheets at the calculated positions on the inner surface of the magnet~\cite{wangPassiveShimmingMethod2022}, using a magnet design with more permanent magnets~\cite{cooleyDesignSparseHalbach2018} and using the focused field approach when designing the magnet array~\cite{rehbergAnalyticApproachCreating2025a}.  To further improve $\mathbf{B}_0$ field homogeneity, active shimming methods can be employed by using the gradient coils and/or dedicated $\mathbf{B}_0$ shim coils~\cite{wuShimCoilDesign2018,stockmann31channelIntegratedAC2022} together with open source software for MR based $\mathbf{B}_0$ shimming~\cite{dastousShimmingToolboxOpensource2023}. Remaining limitations in $\mathbf{B}_0$ field homogeneity and associated image distortions in low-field MRI can also be corrected within the image reconstruction using neural networks ~\cite{schoteJoint$textB_0$Image2024}.\\

\section{Conclusion}\label{conclusions}

Shimmer is an open-source pipeline that allows to effectively improve the $\mathbf{B}_0$ field homogeneity of low-field MRI magnets by an order of magnitude. It has been tested on various magnets and by independent groups and can be adjusted to variable magnet designs. It enables optimization of low-cost inhomogeneous magnets to be utilized for high-quality MR imaging applications. As an open-source project it invites for contributions and extensions to improve its performance and increase its range of applications. 


\section*{Acknowledgments}
The authors would like to thank all the authors that are sharing their work open-source and all the supporters of the Open Source Imaging Initiative (OSI$^2$). The authors want furthermore thank Tom O'Reilly and Victor Kashirin for valuable discussions and contributions. 
The project 22HLT02 A4IM has received funding from the European Partnership on Metrology, co-financed by the European Union's Horizon Europe Research and Innovation Programme and by the Participating States. 
This research is funded by dtec.bw Digitalization and Technology Research Center of the Bundeswehr. dtec.bw is funded by the European Union - NextGeneration EU.

\section*{Financial disclosure}
None reported.

\section*{Conflict of interest}
The authors declare no potential conflict of interests.
\vspace*{-10pt}
\bibliographystyle{elsarticle-num}
\bibliography{shimming_soft_wos}

\newpage

\section*{Supporting information}

Additional supporting information may be found in the
online version of the article at the publisher’s website.

\appendix
\renewcommand{\thesection}{S\arabic{section}}
\setcounter{section}{0}
\renewcommand{\thefigure}{S\arabic{figure}}
\setcounter{figure}{0}

\section{Main magnet simulations}

\label{SI_expected_homos}

\begin{figure}[h]
\centerline{\includegraphics[width=0.75\textwidth]{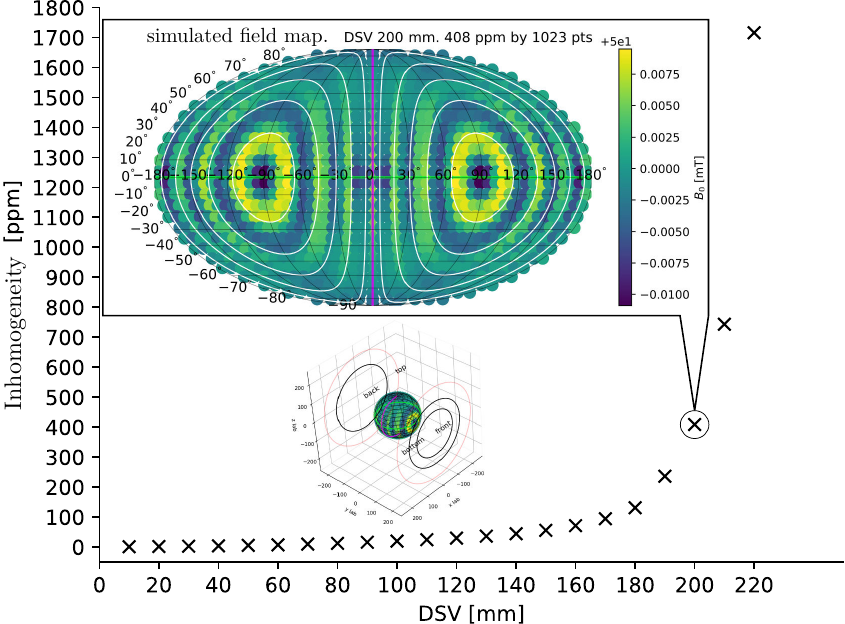}}
\caption{Simulated $\mathbf{B}_0$ homogeneity with respect to the DSV in the OSI$^2$ ONE Low-Field MRI Magnet v2.0. Projections of the rings of the main magnet are highlighted in white. At DSV=200~mm and DSV=240 mm the simulated field inhomogeneities are 408~ppm and 8460 ppm respectively. \label{fig_SI_expected_homos}}
\end{figure}

The distribution of $\mathbf{B}_0$ inside the main magnet is inhomogeneous, especially at the regions that are close to the individual magnets. Figure~\ref{fig_SI_expected_homos} shows the field homogeneity as a function of the DSV in the OSI$^2$ ONE low-field MRI magnet. At 200~mm DSV (inset on the left), the field homogeneity of the ideally assembled magnet should be only 408~ppm, see the simulated field map in the inset in Figure~\ref{fig_SI_expected_homos}.The simulated field map expresses the characteristic concentric-ring pattern that is also seen in the measured field maps at larger DSV.\\

The development of the regularly scaled, concentric circles in the field maps with later shimming iterations can be connected with the equidistantly spaced rings in the main magnet array. The positions of the rings of the main magnet array are shown in white in the field map in Figure~\ref{fig_SI_expected_homos}. This structure of concentric rings is seen in the unshimmed maps and becomes more pronounced as the field is shimmed. The position and the magnitude of this structure does not change upon the shimming iterations, therefore it is concluded that this artifact is inherent to the design of the magnet. Figure~\ref{fig_SI_SHIMMING_PROGRESS_MAGNET_V1}-~\ref{fig_SI_SHIMMING_PROGRESS_MAGNET_V2p0} illustrate the appearance of the ring-like structures due to the equidistantly spaced rings of the main magnet array.

\newpage
\section{Decomposition of the Field Map into Spherical Harmonics\label{SI_SH}}
The distortions of the field map are represented with the multipolar expansion in the basis set of the spherical harmonics (monopole, dipole, quadrupole etc.) The angular dependence of the base functions is $Y_n^m$, whereas the radial distribution is given by $(r/R_0)^{n}$ where $n$ is the order of the corresponding spherical harmonic $Y_n^m$ and $R_0$ is the radius at which the field map was recorded:

\begin{equation}
    B(r,\theta,\phi) \approx \sum_{n=0}^{n_{max}} \sum_{m=-n}^{n} a_{n,m} \left( \frac{r}{R_0}\right)^n Y_n^m\left(\theta,\phi\right)
\end{equation}

\begin{figure}[h]
\centerline{\includegraphics[width=1\textwidth]{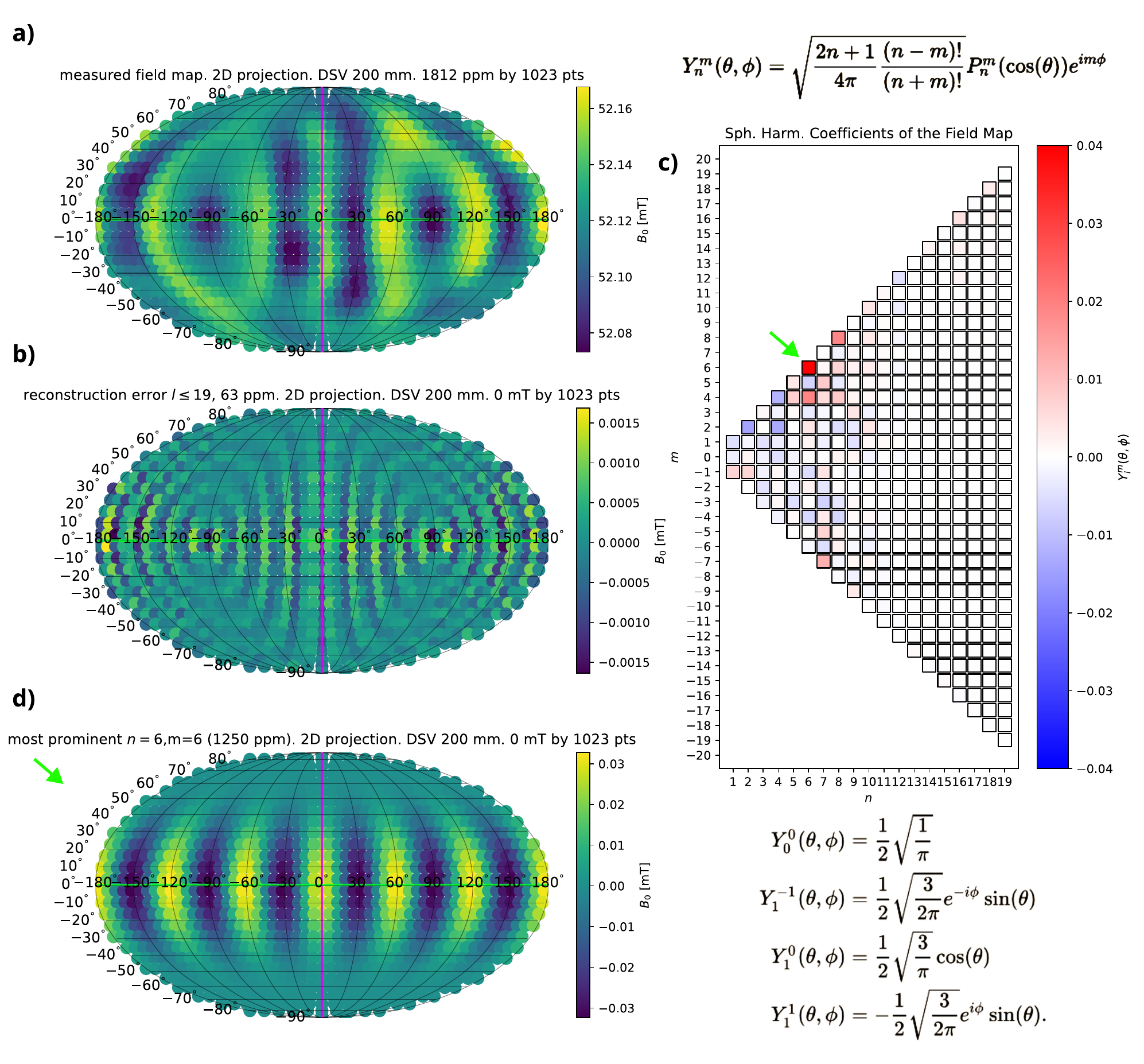}}
\caption{Decomposition of the measured $\mathbf{B}_0$ field map (200~mm DSV) in a discrete set of spherical harmonics $Y_l^m$. a) initial measured map, b) error of the reconstruction of the initial map using SH of orders up to 19. c): SH coefficients that make up the field map, d): The most pronounced mode $n=6, m=6$.\label{fig_SI_SH}}
\end{figure}

Here $a_{n,m}$ are fit parameters. Figure~\ref{fig_SI_SH} shows a decomposition of the measured field map (a, DSV 200~mm) into the first 19 spherical harmonics. The expansion allows for precise reconstruction of the field (b) and one sees the most prominent harmonic in the expansion (c). The angular distribution of the field in the first 5 spherical harmonics is shown in (d).

\clearpage
\section{Performance of the Shimming Pipeline on Three OSI$^2$ ONE Magnets\label{SI_SHIMMING_PROGRESS}}
\begin{figure}[h]
\centerline{\includegraphics[width=0.85\textwidth]{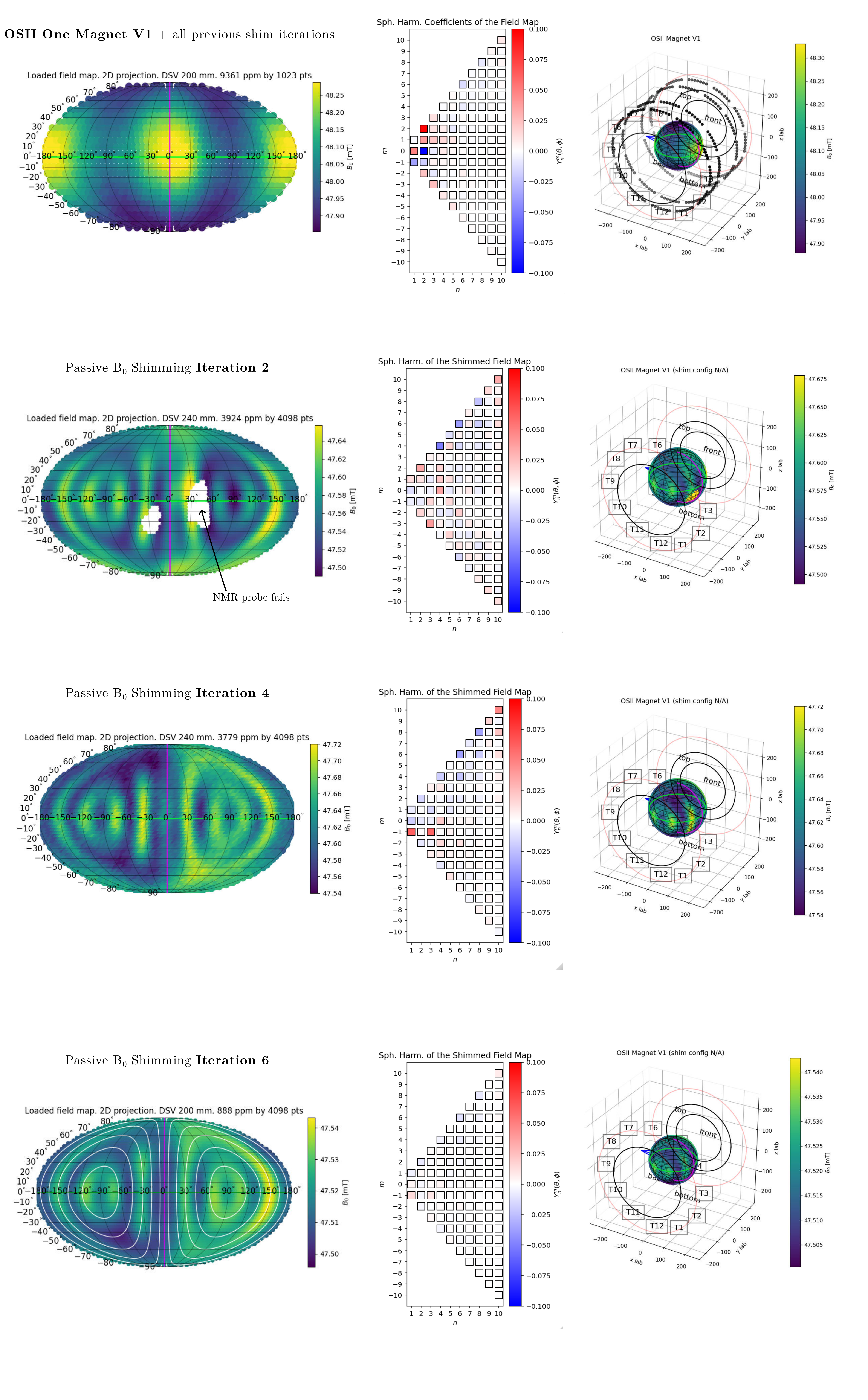}}
\caption{Improvement of the $\mathbf{B}_0$ field map (200-240~mm DSV) upon successive passive shimming iterations in the OSI$^2$ ONE magnet v1.0. Computed ring structures of the main magnet array are shown in white circles on the shimmed field map.  \label{fig_SI_SHIMMING_PROGRESS_MAGNET_V1}}
\end{figure}

The passive $\mathbf{B}_0$ shimming procedure resulted in the improvement of the $\mathbf{B}_0$ homogeneity in three low-field MR magnet arrays: OSI$^2$ ONE magnet v1.0, v2.0 and v2.1. The results of the improvement of the field maps together with the spherical-harmonic decomposition of the field maps are shown in the following figures (Figure~\ref{fig_SI_SHIMMING_PROGRESS_MAGNET_V2p0},Figure~\ref{fig_SI_SHIMMING_PROGRESS_MAGNET_V2p1} and Figure~\ref{fig_SI_SHIMMING_PROGRESS_MAGNET_V1}).

\clearpage

\clearpage
\begin{figure}[h]
\centerline{\includegraphics[width=0.83\textwidth]{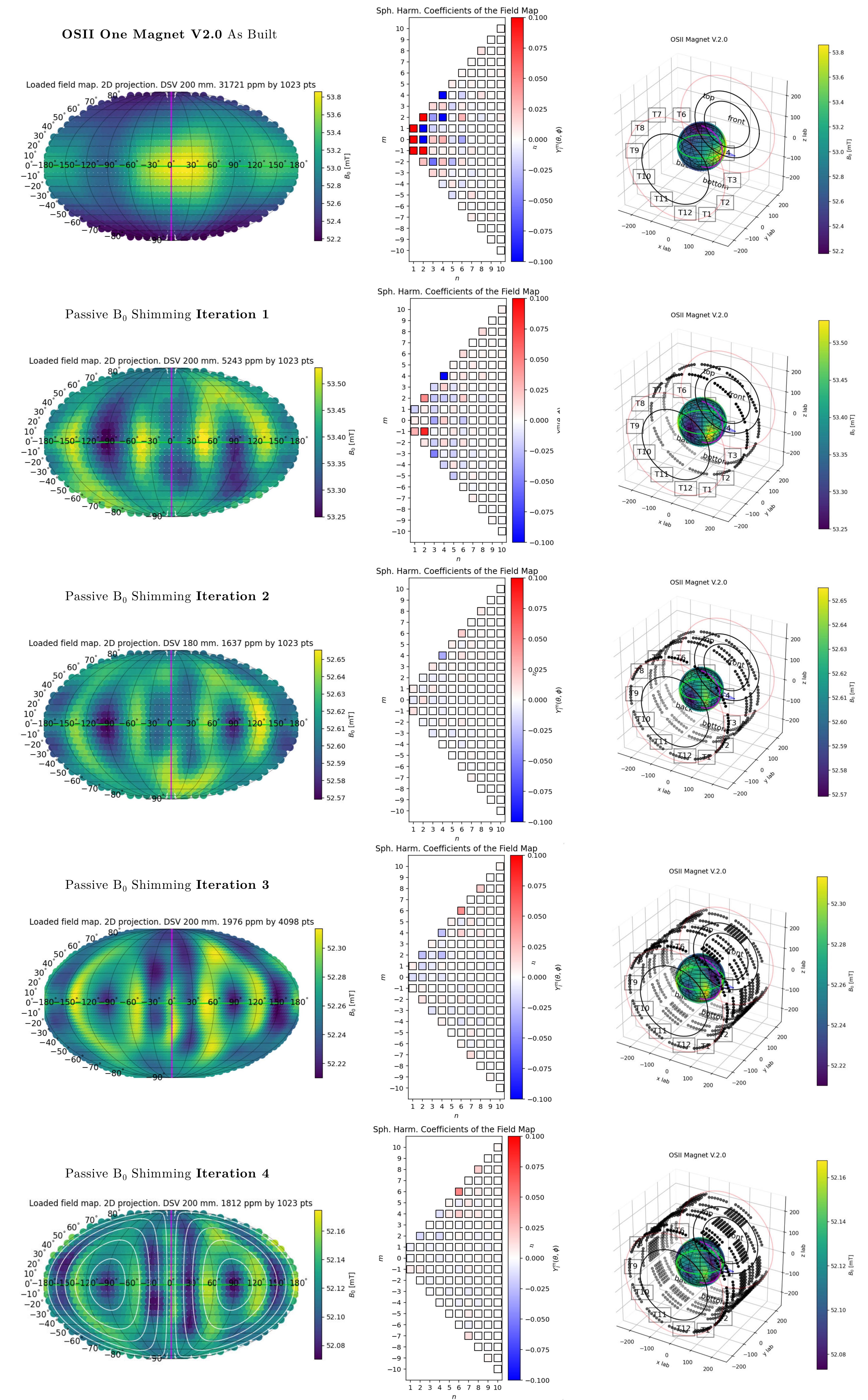}}
\caption{Improvement of the $\mathbf{B}_0$ field map (180-200~mm DSV) upon successive passive shimming iterations in the OSI$^2$ ONE magnet v2.0. Computed ring structures of the main magnet array are shown in white circles on the shimmed field map. \label{fig_SI_SHIMMING_PROGRESS_MAGNET_V2p0}}
\end{figure}

\clearpage
\begin{figure}[h]
\centerline{\includegraphics[width=0.85\textwidth]{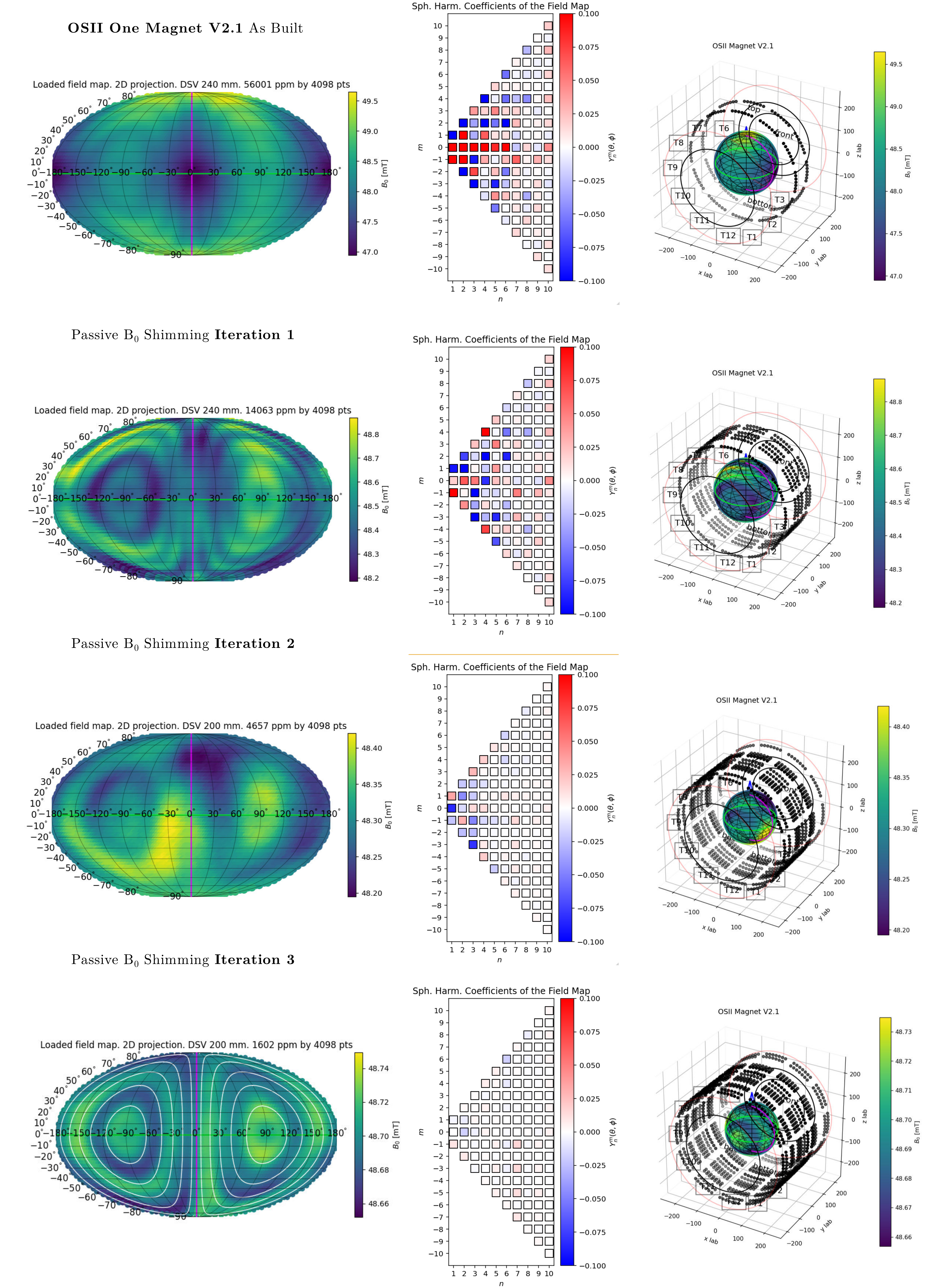}}
\caption{Improvement of the $\mathbf{B}_0$ field map (200-240~mm DSV) upon successive passive shimming iterations in the OSI$^2$ ONE magnet v2.1. Computed ring structures of the main magnet array are shown in white circles on the shimmed field map. \label{fig_SI_SHIMMING_PROGRESS_MAGNET_V2p1}}
\end{figure}

\clearpage
\section{Main magnet uncertainties}\label{SI_error_sources}
\begin{figure}[h]
\centerline{\includegraphics[width=1\textwidth]{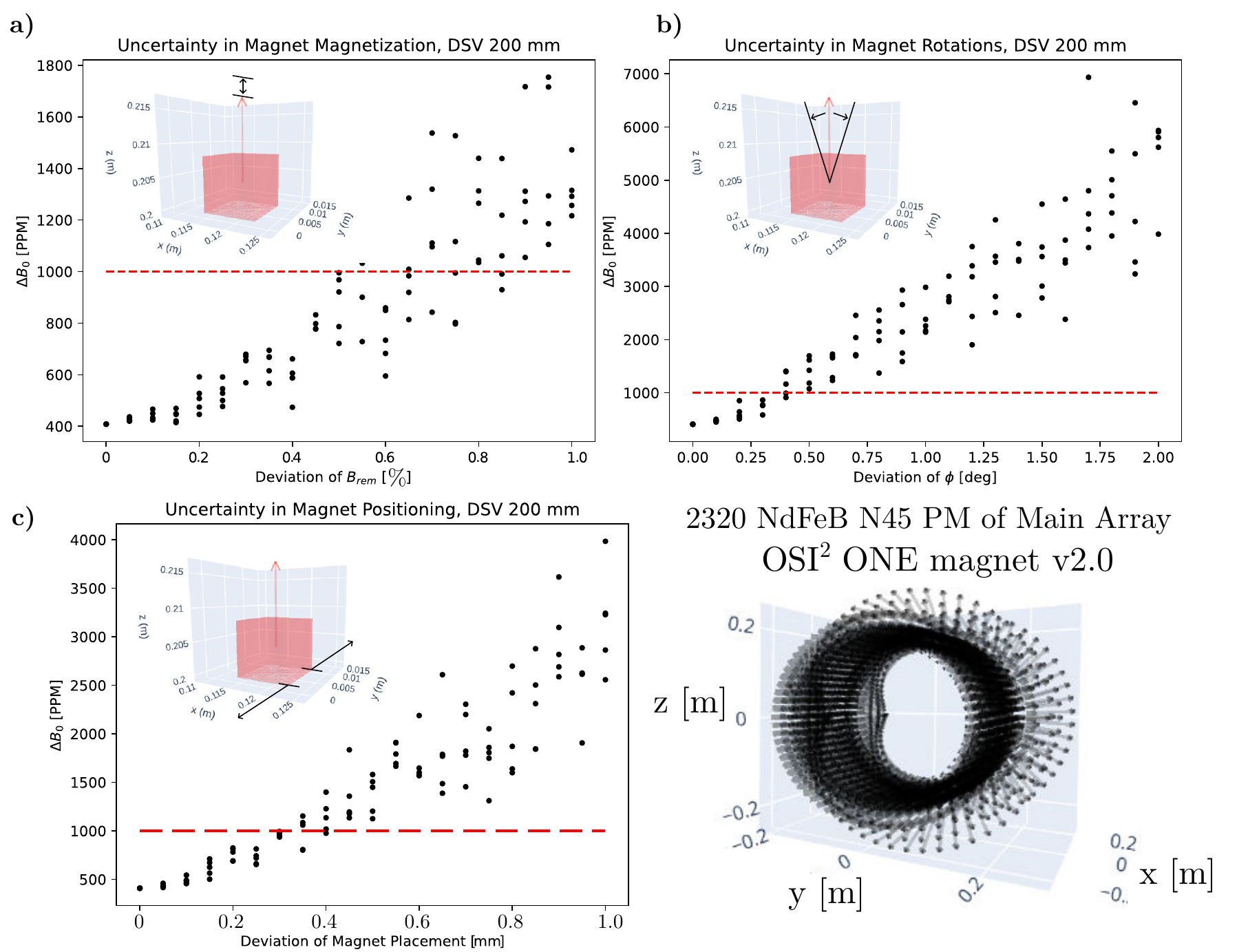}}
\caption{Effect of uncertainties  on the $\mathbf{B}_0$ homogeneity of the main magnet array due to changes in (a), remanence of the magnets, (b) magnet rotations and (c) magnet displacement. The simulations were performed for a 200~mm DSV  of the OSI$^2$ ONE magnet v2.0 magnet. Inset: Illustrated parameter deviation applied in the simulation.\label{fig_SI_delta_phi}\label{fig_SI_delta_y}\label{fig_SI_delta_brem}}
\end{figure}

The error in the field of the assembled magnet comes mostly from the uncertainties in the placement of the permanent magnets and to a smaller extent from the uncertainties in the magnet properties. Figure~\ref{fig_SI_delta_phi} demonstrates the field deviation due to uncertainties in magnet placement, magnet rotations and magnet magnetization for OSI$^2$ ONE Magnet v2.0. The values of the parameters were changed according to the normal (Gaussian) distribution. The x axes of the corresponding plots refer to the standard deviation ($\sigma$ for the corresponding distributions). The field maps were rendered on a 200~mm DSV at 1024 equidistantly spaced points. The following plots set the requirements on the assembly quality. For example, to achieve a 1000~ppm field inhomogeneity in a 200~mm DSV: $\Delta \phi\leq$0.3$^\circ$, $\Delta x\leq$~0.3mm, $\Delta B_{rem}\leq$0.6~\%.

\section{Shim magnet uncertainties}\label{SI_shim_magnet_uncertainties}
\begin{figure}[h]
\centerline{\includegraphics[width=1\textwidth]{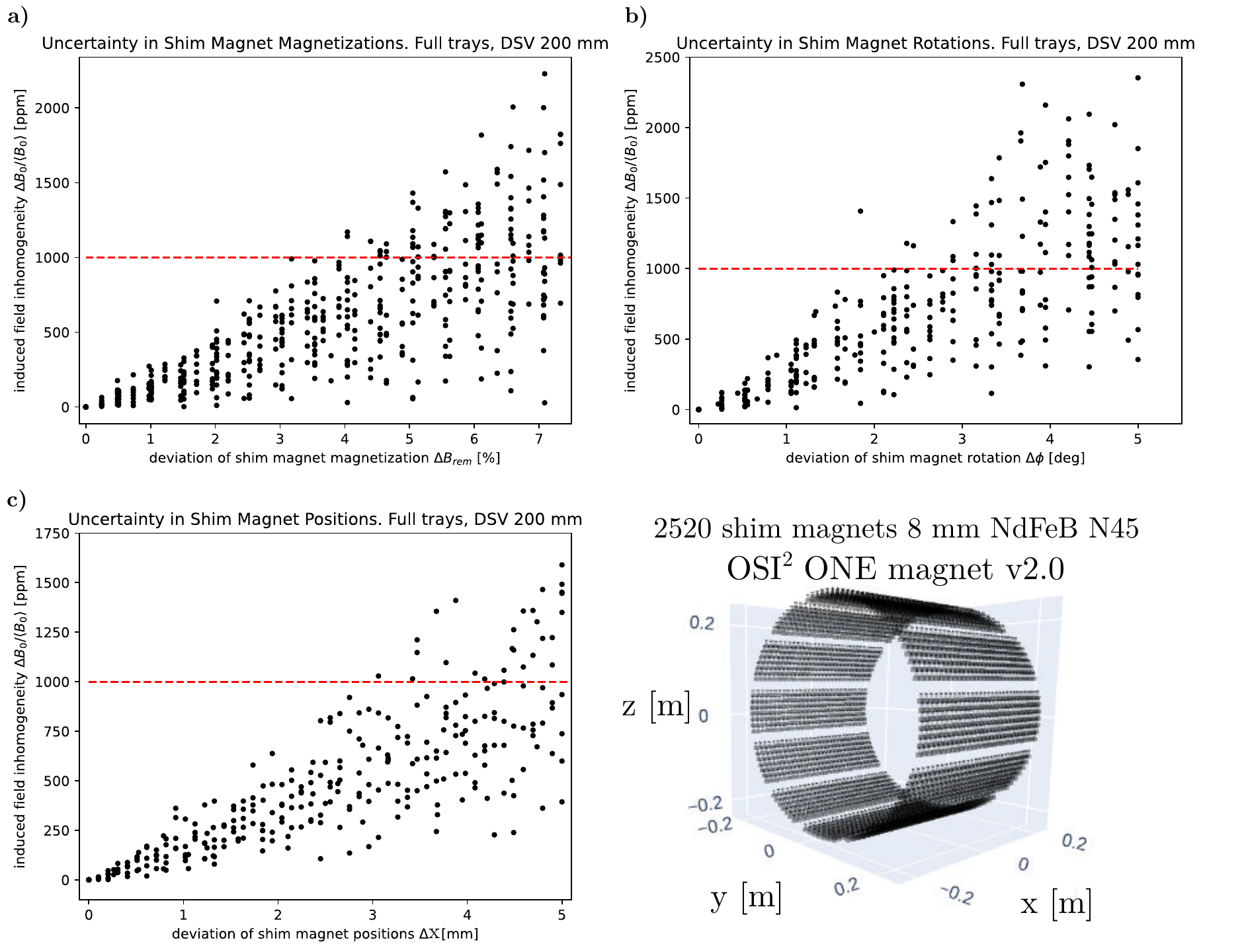}}
\caption{Effect of uncertainties on the $\mathbf{B}_0$ homogeneity due to changes in (a), remanence of the shim magnets, (b) shim magnet rotations and (c) shim magnet displacement. \label{fig_SI_shim_magnet_uncertainty}}
\end{figure}

The shim magnets are placed further away from the DSV so it is expected that the errors in their positioning, magnetization and rotations will have lighter effect on the field inhomogeneity as compared to the main magnet array. Figure~\ref{fig_SI_shim_magnet_uncertainty} illustrates the field deviation due to uncertainties in shim magnet placement, shim magnet rotations and shim magnet magnetization for OSI$^2$ ONE Magnet v2.0. To compute the inhomogeneity increase for each parameters (deviation of angle of magnets, deviation of position and deviation of magnetization), a full shim configuration was manipulated accordingly. The values of the parameters were changed according to the normal (Gaussian) distribution. The x axes of the corresponding plots refer to the standard deviation ($\sigma$ for the corresponding distributions). The shim fields were rendered on a 200-mm DSV at 1024 equidistantly spaced points. The following plots set the requirements on the assembly quality for the shim trays. For example, to have a deviation of 1000~ppm in field inhomogeneity in a 200~mm DSV: $\Delta \phi\leq$2$^\circ$, $\Delta x\leq$3~mm, $\Delta B_{rem}\leq$3~\%.

\end{document}